\documentclass[12pt]{article}
\usepackage{graphicx}
\usepackage{amsfonts}
\usepackage{amssymb,amsmath}
\usepackage{latexsym}
\usepackage{color}
\usepackage{cite}
\usepackage{bm}
\usepackage{ulem}

\input{colordvi.tex}

\begin{document}

\setcounter{page}{1}

\begin{titlepage}
\begin{center}

\vskip .5in

{\Large \bf Constraining light gravitino mass with 21 cm line observation}

\vskip .45in

{\large
Yoshihiko~Oyama$\,^1$ and 
Masahiro~Kawasaki$\,^{1}$
}

\vskip .45in

{\it
$^1$
Institute for Cosmic Ray Research, The University of Tokyo, 5-1-5 Kashiwanoha, Kashiwa, Chiba 277-8582, Japan \vspace{2mm} 
}

\end{center}

\vskip .4in

\begin{abstract}
We investigate how well we can constrain 
the mass of light gravitino $m_{3/2}$
by using future observations of 21 cm line fluctuations
such as Square Kilometre Array~(SKA) and Omniscope.
Models with light gravitino with the mass $m_{3/2} \lesssim \mathcal{O}(10)$~eV 
are quite interesting 
because they are free from the  cosmological gravitino problem
and consistent with many baryogenesis/leptogenesis scenarios.
We evaluate expected constraints on the mass of light gravitino
from the observations of 21 cm line,
and show that the observations are
quite useful to prove the mass.
If the gravitino mass is $m_{3/2}=1$~eV,
we found expected 1~$\sigma$ errors on $m_{3/2}$ are
$\sigma(m_{3/2})=0.25$~eV (SKA phase~1),
$0.16$~eV (SKA phase~2) and $0.067$~eV (Omniscope)
in combination with Planck + Simons Array + DESI~(BAO) + $H_0$. 
Additionally, we also discuss 
detectability of the effective number of neutrino species 
by varying the effective number of neutrino species for light gravitino $N_{3/2}$
and constraints on the mass of light gravitino in the presence of massive neutrinos.
We show that 21 cm line observations 
can detect the nonzero value of $N_{3/2}$, and allow us to distinguish 
the effects of the light gravitino from those of massive neutrino.

\end{abstract}
\end{titlepage}

\setcounter{footnote}{0}

\section{Introduction}
\label{sec:Intro}

In particle physics models with  local super symmetry~(SUSY) or supergravity,
one of the most important predictions
is the existing of gravitino, which is the superpartner of graviton and has a spin 3/2.
The gravitino mass $m_{3/2}$
is related to the energy scales of SUSY breaking,
and can vary from an order of  eV up to of TeV.
In particular, scenarios with 
light gravitino whose mass is $m_{3/2} \lesssim \mathcal{O}(10)$~eV 
are very interesting 
because they are free from the  cosmological gravitino problem~\cite{Moroi:1993mb},
and can be
consistent with some baryogenesis scenarios 
which require  a high reheating temperature such as thermal leptogenesis~\cite{Fukugita:1986hr}.
Therefore, the scenarios with the light gravitino
are very attractive in cosmology.

It is important to determine the gravitino mass
in order to understand the mechanism of SUSY breaking.
Although it can be probed by collider experiments~(e.g. LHC)~\cite{Hamaguchi:2007ge}
with direct and indirect signatures,
we can also obtain constraints on it
from cosmological observations~\cite{Pierpaoli:1997im}.
In the early Universe,
light gravitinos are produced from thermal plasma,
and they behave as warm dark matter~(WDM)  at late epochs.
The light gravitinos influence the growth of density fluctuations 
mainly through the following two effects.
First of all, they change the time of matter-radiation equality
because the light gravitinos behave as a radiation component at early epochs.
Secondly,  the light gravitinos have large velocity dispersions
and propagate up 
to the horizon scales until they become non-relativistic.
%
Then, they erase their own density fluctuation
and suppress the growth of density fluctuations of matter
below the free-streaming scale.
However, the former effect is very small
because the energy density of light gravitinos
does not have a large  fraction of the total energy of radiation. 
Therefore, 
we can probe signatures of the light gravitino mainly
through the latter effect.
From Lyman-$\alpha$ forest data in combination with WMAP \cite{Viel:2005qj},
a constraint on the light gravitino mass is obtained, 
and its bound is $m_{3/2}<16$~eV (95\% C.L.).
Additionally, some authors have pointed out that
measurements of CMB lensing~\cite{Ichikawa:2009ir} 
or weak lensing surveys of galaxies~\cite{Kamada:2013sya}
are quite effective in constraining the light gravitino mass.
By using the date of CMB lensing from Planck and cosmic shear from the CHFTLenS survey,
a stringent constraint is obtained,
and the upper bound is $m_{3/2}<4.7$~eV (95\% C.L.)~\cite{Osato:2016ixc}.
However, 
it is difficult to obtain significant bounds of the light gravitino mass
if we treat the effective number of neutrino species 
for light gravitino 
$N_{3/2}\equiv \rho_{3/2}/\rho_{\nu}$ as a free parameter, 
where $\rho_{3/2}$ and $\rho_{\nu}$
are the energy densities of light  relativistic gravitinos and neutrinos, respectively.
Moreover,
discriminating signatures of light gravitino from
that of massive neutrino is also quite difficult.
%
Therefore, if one wants to discriminate light gravitinos from other possibilities, 
it is mandatory to find a new powerful probe of cosmological signatures of the light gravitino.

In this paper,  we particularly investigate 
the issue of how accurately we can constrain the light gravitino mass
by using future observations of fluctuations of neutral hydrogen 21 cm line
which comes from the epoch of reionization (EoR), in addition to those of CMB.
By observing the power spectrum of cosmological 21 cm line fluctuations, 
we will be able to obtain useful information on 
a variety of cosmological parameters
\cite{McQuinn:2005hk,Loeb:2008hg,Pritchard:2008wy,
Pritchard:2009zz,Abazajian:2011dt,Oyama:2012tq,Kohri:2013mxa,Kohri:2014hea,Oyama:2015vva,Oyama:2015gma}.
%
%
%
%
%
%
Because observations of the 21 cm line can cover a wide redshift range, 
they can be complementary to other observations such as CMB. 
Additionally, 
the effects of the light gravitino mainly appear on small scales, 
which can be well measured by 21 cm observations. 
In order to discuss expected constraints 
from the future cosmological surveys 
on the mass of light gravitino,
we make Fisher analysis by assuming the specifications 
for planned observations of 21 cm line 
such as Square Kilometre Array (SKA) \cite{SKA} and Omniscope~\cite{Tegmark:2009kv,Tegmark:2008au}. 
In our analysis, 
we also take into account 
future CMB observations 
such as the Simons Array~\cite{SimonsArray} and COrE+~\cite{COrE+}. 
Besides, we consider
including information of a baryon acoustic oscillation (BAO) observation,
such as Dark Energy Spectroscopic Instrument (DESI) \cite{DESI:web}
and a direct measurement of the Hubble constant $H_0$.

This paper is organized as follows. 
In Section~\ref{sec:Light-gravi}, 
we briefly review the effects of light gravitino on cosmology.
In Section~\ref{sec:methods}, 
we review analytical methods used in this paper,
paying particular attention to
21 cm line, CMB, BAO observations and the direct measurement of $H_0$.
We show our results in Section~\ref{sec:results}, 
and Section~\ref{sec:conclusion} is devoted to our conclusion.

\section{Light gravitino and its effects on large-scale structure in the Universe}
\label{sec:Light-gravi}

\subsection{Light gravitino}
\label{subsec:Light-gravi}

The existence of light gravitino is typically predicted in 
gauge-mediated SUSY breaking (GMSB) 
scenarios \cite{Dine:1981gu, Nappi:1982hm, AlvarezGaume:1981wy, 
Dine:1993yw, Dine:1994vc, Dine:1995ag}.
%
SUSY breaking is the origin of 
the masses of the gravitino and SUSY particles.
The SUSY breaking field $S$
has a vacuum expectation value as $\left< S\right> = M + F_S \theta^2$,
and $F_S$ gives SUSY breaking scale,
which is related to the gravitino mass $m_{3/2}$.
The gravitino mass is given by
\begin{align}
m_{3/2} = \frac{F_S}{\sqrt{3} M_{\textrm{pl}}},
\end{align}
where $M_{\textrm{pl}} \simeq 2.4\times 10^{18}$ GeV is the reduced Planck mass.
%
On the other hand, in GMSB scenarios,
sparticles in the standard model~(SM) sector  acquire their masses
through messenger fields, whose mass scale is denoted as $M_{\textrm{mess}}$.
%
For example, in a GMSB model with
$N$ pairs of messenger particles,
gaugino masses $M_a$ 
($a=1,2,3$ is a gauge group index)
and sfermion masses squared $m_{\tilde{f}_i}^2$
($i$ is a flavor index) are typically given by
\begin{align}
M_{a} & = 
N \left( \frac{\alpha_{a}}{4\pi}\right)
\frac{F_S}{M_{\textrm{mess}}}, \\
m_{\tilde{f}_i}^2 & =
2N \sum_a C_a^{(i)} 
\left( \frac{\alpha_{a}}{4\pi}\right)^2
\left(\frac{F_S}{M_{\textrm{mess}}} \right)^2,
\end{align}
where 
$C_a^{(i)}$ is Casimir operators for the sfermion $\tilde{f}_i$,
and $\alpha_a$ denotes the gauge coupling constants.
%
Although 
$(F_S/M_{\textrm{mess}})\sim 100$~TeV is required
in order to obtain TeV scale masses,
still the SUSY breaking scale $F_S$ or the gravitino mass $m_{3/2}$
can take a wide range of values as 
$\mathcal{O}(1)\textrm{eV}\lesssim m_{3/2}\lesssim \mathcal{O}(10)\textrm{GeV}$,
where the upper bound comes from the requirement that
the gravity-mediation does not dominate,
and existence of the lower bound 
arises from avoiding destabilization of the messenger scalar 
and not leading to unwanted vacua.


In the GMSB models,
stringent constraints on the gravitino mass
are obtained from the Higgs mass measured by LHC~\cite{Aad:2015zhl,Chatrchyan:2012xdj}.
In the minimal supersymmetric standard model (MSSM),
the stop mass is required to be as large as $\mathcal{O}(10-100)$~TeV
in order to achieve the measured large Higgs mass $m_h=125$~GeV,
and the bound can place a lower bound on the gravitino mass.
In a class of GMSB models with $N$ copies of messenger fields in the
$5+\bar{5}$ representation of SU(5),
we can obtain a bound
$300~\textrm{eV} < m_{3/2}$ with $N=1$, and 
$60~\textrm{eV} < m_{3/2}$ with $N=5$~\cite{Ajaib:2012vc},
if the coupling between the messengers and the SUSY breaking field is perturbative.
Although a range of the light gravitino mass 
which is allowed by present cosmological observations 
are ruled out,
the bound by the LHC is model-dependent.
For example,
$\mathcal{O}(1-10)$~eV may be possible 
if the coupling is non-perturbative or
a singlet Higgs is introduced (next to MSSM) \cite{Yanagida:2012ef}.

Additionally,
less stringent lower bounds are obtained
from direct SUSY searches in LHC~\cite{Aad:2014mra}.
Assuming a perturbative coupling,
we can obtain a lower bound 
$3.7 \textrm{eV} < m_{3/2}$
in the same GMSB model as mentioned above
with $M_{\textrm{mess}}=250$~TeV and $N=3$ ($10+\bar{10}$ of SU(5)).

\subsection{Effects of light gravitino on the growth of density fluctuations}
\label{subsec:Light-gravi_effects}

If we take account of the presence of the light gravitino in the early Universe,
some difficulties arise in constructing a consistent cosmological scenarios, 
which we are going to describe shortly below.
At the reheating era, gravitinos are efficiently produced
and the abundance of them can easily exceed 
that of the present dark matter
unless the reheating temperature  $T_R$ is very low~\cite{Moroi:1993mb}.
However, many known baryogenesis/leptogenesis scenarios
require high enough reheating temperature,
which may be inconsistent with 
the upper bound coming from the observed abundance of dark matter.
For example, thermal leptogenesis scenario~\cite{Fukugita:1986hr} 
requires $T_{\rm R}\gtrsim 10^{9}$~GeV  
and the reheating temperature 
seems to conflict with the gravitino problem
except for the very light gravitino mass range 
$m_{3/2}\lesssim 100$~eV.
If gravitinos have such a small mass,
they are thermalized in the early Universe~\cite{Ichikawa:2009ir,Pierpaoli:1997im}.
In that case,
their abundance does not have dependency on 
the reheating temperature
and is smaller than the dark matter density 
if $m_{3/2}\lesssim 100$~eV.
This advantage is the reason why
we particularly focus on the light gravitino scenario.

Thermally produced gravitinos 
decouple from the other particles at some point,
and their relic abundance is fixed.
The number density is determined by 
the effective degrees 
of freedom 
$g_{*3/2}$ at the decoupling of the gravitinos.
In \cite{Ichikawa:2009ir,Pierpaoli:1997im}, the number density of the gravitinos 
in GMSB models is evaluated by solving the Boltzmann equation,
and for a messenger mass scale $M_{\textrm{mess}}\sim 100$~TeV,
$g_{*3/2}$ becomes $g_{*3/2}\sim 90$~\cite{Ichikawa:2009ir}
with only mild dependence on $m_{3/2}$.
In consideration of the result,
we set $g_{*3/2} = 90$ as the fiducial value
in our analysis.

The thermally produced light gravitinos
behave as a warm dark matter component~\cite{Ichikawa:2009ir,Pierpaoli:1997im},
and they can be parametrized by their temperature and mass.
Since gravitinos interact with other particles
through their goldstino components in the GMSB scenario,
their phase-space distribution
is a Fermi-Dirac distribution with two degrees of freedom.
Because light gravitinos behave as relativistic particles 
at early epoch,
we can parametrize its energy density $\rho_{3/2}$ 
by using the effective number of neutrino species,
and it is given by
\begin{align}
N_{3/2}
=\frac{\rho_{3/2}}{\rho_{\nu}}=\left(\frac{T_{3/2}}{T_{\nu}}\right)^{4}
=\left(\frac{g_{*\nu}}{g_{*3/2}}\right)^{4/3},
\label{eq:effective_num_gravi}
\end{align}
where $\rho_{\nu}$ is the energy density of one generation of neutrinos , 
and $g_{*3/2}$ and $g_{*\nu}$ are the effective degrees of freedom at decoupling of
light gravitinos and neutrinos, respectively.
In standard cosmology, 
the degree of freedom of neutrinos at the neutrino decoupling
is $g_{*\nu}=10.75$.
$T_{3/2}$ and $T_{\nu}$ are temperatures of 
light gravitinos and neutrinos, respectively.
From Eq.~\eqref{eq:effective_num_gravi}, 
the temperature of light gravitino at present is evaluated  as
\begin{align}
T_{3/2}=\left( N_{3/2}\right)^{1/4}T_{\nu}=1.95\left( N_{3/2}\right)^{1/4} \textrm{K},
\label{eq:temp_gravi}
\end{align}
where we use the temperature of neutrinos in the standard cosmology.

At late epochs, light gravitinos lose their energy and
become non-relativistic particles due to the cosmic expansion.
Its present density parameter $\Omega_{3/2}h^2$ can be estimated as
\begin{align}
\Omega_{3/2}h^2=0.1269 \left(\frac{m_{3/2}}{100~\textrm{eV}}\right)\left(\frac{90}{g_{*3/2}}\right).
\label{eq:Omega_3_2}
\end{align}
In the following, 
we assume that dark matter does not consists solely of the light gravitinos 
because the gravitino mass $m_{3/2}$ needs to be about 90 eV in order to 
be consistent with  observed dark matter density $\Omega_{\textrm{DM}}\simeq 0.11$~\cite{Ade:2015xua},
which contradicts with
the constraint from Ly-$\alpha$ forest, $m_{3/2}<16$~eV~\cite{Viel:2005qj},
as well as a more recent one
$m_{3/2}<4.7$~eV from CMB and cosmic shear data~\cite{Osato:2016ixc}.
Therefore, we assume that dark matter consists of
the light gravitino and CDM components.
i.e. $\Omega_{\textrm{DM}}=\Omega_{\textrm{CDM}}+\Omega_{3/2}$,
and  we define the fraction of gravitino 
in the total dark matter density as
\begin{align}
f_{3/2}\equiv \frac{\Omega_{3/2}}{\Omega_{\textrm{DM}}}.
\end{align}
As the CDM component,
the QCD axion, a messenger baryon proposed in~\cite{Hamaguchi:2007rb}
and so on can be well-fitted 
within the framework of the GMSB scenario.

From now on, we briefly explain effects of the light gravitinos on cosmological structure formation.
They behave as a warm dark matter component, 
and their effects on structure formation can be understood by 
considering the following two aspects:
(i) a contribution to the energy density of radiation 
(ii) suppression of matter fluctuations on small scales through the free-streaming. 
The first effect is due to the fact 
that light gravitinos behave as  relativistic particles at early epochs.
Therefore, the time of matter-radiation equality is slightly delayed 
if light gravitinos exist in the Universe.
The delay alters the evolution of gravitational potential 
and drives the integrated Sachs-Wolfe~(ISW) effect
in the CMB temperature anisotropy.
In addition, the matter fluctuations are suppressed at small scales 
through stagspansion effect due to the delaying of matter-radiation equality.
However, its contribution is so small 
(as we mentioned before, 
the theoretical calculation predicts 
that $g_{*3/2}$ is around 90, which corresponds to $N_{3/2}\simeq0.059$~\cite{Ichikawa:2009ir}),
and it is difficult to measure the impacts due to this effect
by observing 
CMB anisotropies without lensing.
Therefore, constraints on the gravitino mass 
mainly come from the second effect, 
i.e. its free-streaming behavior.
Because light gravitinos have relatively large thermal velocity, 
they 
propagate up to their free-streaming scale 
and erase own density fluctuation
in a similar manner to massive neutrinos.
Within the free-streaming scale,
light gravitinos do not contribute to the gravitational growth of the matter fluctuations.
Thus,  matter fluctuations at small scales are suppressed in comparison to
the $\Lambda$CDM model.

Massive neutrinos also have similar effects on the growth of matter fluctuations,
but its temperature and energy density are different from those of light gravitinos.
Therefore, in principle, we can discriminate between
the effects of the light gravitino and the massive neutrino
through observing their free streaming scale.

\section{Forecasting methods}
\label{sec:methods}

\subsection{21 cm  line}

Here, we briefly review a forecasting method 
related to 21 cm line observations in our analysis.
For further details of the 21 cm line observations, 
we refer to Refs.~\cite{Furlanetto:2006jb,Pritchard:2011xb}.

\subsubsection{Power spectrum of 21 cm  radiation}

The 21 cm line of neutral hydrogen atom is emitted by transition
between the hyperfine splitting of the 1s ground state. 
%
We can observe signals of 21 cm line which come from the epoch of reionization~(EoR)
or the cosmic dark ages
as the differential brightness temperature 
relative to the temperature of CMB $T_{\rm CMB}$:
\begin{align}
\Delta T_{b} \left(\mbox{\boldmath $r$},z \right) 
&=&
\frac{3c^{3}hA_{21}}{32\pi k_{B}\nu_{21}^{2}}
\frac
{x_{HI}(\mbox{\boldmath $r$},z)n_{H}(\mbox{\boldmath $r$},z) } {(1+z)H(z)}
\left(
1-\frac{T_{{\rm CMB}}(\mbox{\boldmath $r$},z) }{T_{S}(\mbox{\boldmath $r$},z)}
\right)
\left(
1-\frac{1+z}{H(z)}
\frac{dv_{p\|}(\mbox{\boldmath $r$},z)}{dr_{\|}}
\right), 
\label{eq:obsbrightness2}
\end{align}
where 
$\mbox{\boldmath $r$}$ is the comoving coordinates of the source of 21 cm line, 
$z$ represents the redshift at emission/absorption, 
$A_{21} \simeq 2.869 \times 10^{-15}{\rm s^{-1}}$ 
is the spontaneous decay rate of the hyperfine splitting,
$\nu_{21} \simeq 1.42$ GHz is the frequency of 21 cm line,
$n_{H}$ is the number density of hydrogen,
$x_{HI}$ is the fraction of neutral hydrogen,
and $dv_{p\parallel} / dr_{\parallel}$ is the gradient of peculiar velocity 
along the line of sight.
$T_S$ is the spin temperature, which is
defined by $n_1/ n_0 = 3 \exp ( - T_{21} / T_S)$,
where $n_0$ and $n_1$ are the number densities of singlet
and triplet states of neutral hydrogen atom, respectively.
Here $T_{21}= hc / k_B
\lambda_{21}$ is the temperature corresponding to 21 cm line,
and $\lambda_{21}$ is its wavelength.

In this paper, we assume that $T_S \gg T_{\rm CMB}$ 
because we focus on the epoch of reionization
during which this condition is well satisfied.
In general, the brightness temperature is sensitive 
to details of inter-galactic medium (IGM)
and astrophysical processes.
However, with a few reasonable assumptions, 
we can eliminate the dependence from the 21 cm line 
brightness temperature~\cite{Madau:1996cs,Furlanetto:2006tf,Pritchard:2008da}.
At the epoch of reionization long after star formation begins,
X-ray background produced by early stellar remnants heats the IGM.
Therefore, kinetic temperature of the IGM $T_{K}$ becomes much higher than 
that of CMB $T_{\textrm{CMB}}$.
Furthermore, the star formation produces 
a large amount of Ly$\alpha$ photons sufficient to couple
$T_{S}$ to $T_{K}$ through the Wouthuysen-Field effect~\cite{Wouthuysen:1952,Field:1958}.
In this scenario, 
$T_{\textrm{CMB}} \ll T_{K} \sim T_{S} $ are justified at $z \lesssim 10$, 
and $\Delta T_{b}$ does not depend on $T_{S}$.

Now, let us consider fluctuations of 
the differential brightness temperature of 21 cm line $\Delta T_b ({\bm r})$. 
By expanding the hydrogen number density $n_H$ 
and the ionization fraction $x_i$ ($x_i=1-x_{HI}$) as 
$n_H(\bm{r}) = \bar{n}_H (1 + \delta (\bm{r}))$ 
and $x_i(\bm{r}) = \bar{x}_i (1 + \delta_x (\bm{r}) )$, 
we can rewrite Eq.~\eqref{eq:obsbrightness2} as 
\begin{align}
\Delta T_b (\bm{r},z)
=  
\Delta \bar{T}_b(z) \left( 1 - \bar{x}_i ( 1+ \delta_x (\bm{r},z) ) \right) 
(1+ \delta (\bm{r},z) )
\left(
1-\frac{1+z}{H(z)}
\frac{dv_{p\|}(\mbox{\boldmath $r$},z)}{dr_{\|}}
\right),
\end{align}
where we assume that $T_{\textrm{CMB}} \ll T_S $
and neglect the term including the spin temperature.
Here, $\Delta\bar{T}_b$ is the spatially averaged 
differential brightness temperature at redshift $z$ and given by
\begin{align}
\Delta \bar{T}_b(z)
\simeq 26.8 
\left(\frac{1-Y_p}{1-0.25} \right)
\left( \frac{\Omega_b h^2}{0.023} \right)
\left( \frac{0.15}{\Omega_m h^2} \frac{1+z}{10} \right)^{1/2}~\textrm{mK},
\end{align}
where $Y_p$ is the primordial $^4$He mass fraction.
By denoting the fluctuation of $\Delta T_b$ as 
$\delta (\Delta T_b (\bm{r},z)) \equiv \Delta T_b(\bm{r},z) - \bar{x}_{H}(z)\Delta \bar{T}_b$(z),
the 21 cm line power
spectrum $P_{21} (\bm{k})$ in the $k$-space is defined by
\begin{align}
\left\langle
\delta (\Delta T^\ast_b ({\bm k})) \delta (\Delta T_b (\bm{k}'))  \right\rangle
= (2\pi)^3 \delta^3 ( \bm{k-k'}) P_{21} (\bm{k}).
\label{eq:power}
\end{align}
Because the Fourier component of
the peculiar velocity term
$\delta_v \equiv (1+z)(dv_{p\|} / dr_{\|})/H(z)$
is given by
$ \delta_v  ({\bm k})= -\mu^2 \delta  ({\bm k})$ 
within the linear perturbation theory
($\mu = \hat{\bm k}\cdot \hat{\bm n}$
is the cosine of the angle between the wave vector and the line of sight), 
the power spectrum can be written as
\begin{align}
P_{21} (\bm{k}) 
=
P_{\mu^0} (k) + \mu^2 P_{\mu^2} (k)  + \mu^4 P_{\mu^4} (k), 
\end{align}
where  $k = |\bm{k}|$ and
\begin{align}
P_{\mu^0}  & = \mathcal{P}_{\delta \delta} - 2 \mathcal{P}_{x\delta} + \mathcal{P}_{xx}, 
\\
P_{\mu^2}   & = 2 \left( \mathcal{P}_{\delta \delta} - \mathcal{P}_{x\delta} \right),  
\\
P_{\mu^4}  & =   \mathcal{P}_{\delta \delta}. 
\end{align}
Here, $ \mathcal{P}_{\delta \delta} \equiv (\Delta \bar{T}_b)^2 \bar x_{HI}^2P_{\delta \delta}, 
\mathcal{P}_{x \delta} \equiv (\Delta \bar{T}_b)^2 \bar{x}_i \bar{x}_{HI} 
P_{x \delta} $ and $ \mathcal{P}_{x x} \equiv (\Delta \bar{T}_b)^2 \bar{x}_i^2 P_{ x x} $,
where $P_{\delta\delta}, P_{x\delta}$ and $P_{xx}$ 
are the auto- and cross- power spectra 
defined in the same manner as Eq.~\eqref{eq:power} 
for the fluctuation of hydrogen number density $\delta$ 
and that of ionization fraction $\delta_x$.
Since $P_{\delta\delta}$ traces the fluctuation of matter,
the power spectrum of 21 cm line 
has information on cosmological parameters.

$P_{x\delta}$ and $P_{xx}$ can be neglected 
as long as we consider eras when the IGM is completely neutral. 
However, after the reionization starts,
these two spectra significantly contribute 
to the 21 cm line power spectrum.  
In order to evaluate these spectra,
we adopt the treatment given in Ref.~\cite{Mao:2008ug}, 
where they assumed that 
$\mathcal{P}_{x\delta}$ and $\mathcal{P}_{xx}$ have specific forms
which match simulations incorporating radiative transfer 
in Refs.~\cite{McQuinn:2006et,McQuinn:2007dy}.  
The explicit forms of the power spectra are parametrized to be
\begin{align}
\mathcal{P}_{x x}  (k) 
& =
b_{xx}^2 \left[ 1 + \alpha_{xx} (k R_{xx}) + (k R_{xx})^2 \right]^{-\gamma_{xx} / 2} 
\mathcal{P}_{\delta\delta} (k), \label{eq:Pxx}
\\
\mathcal{P}_{x \delta} (k) 
& = 
b_{x\delta}^2 ~e^{ - \alpha_{x\delta} (k R_{x\delta}) - (k R_{x\delta})^2} 
\mathcal{P}_{\delta\delta} (k),
\label{eq:Pxdelta}
\end{align}
where $b_{xx}$, $b_{x\delta}$, $\alpha_{xx}$, $\gamma_{xx}$ and $\alpha_{x\delta}$
are parameters which characterize the amplitudes and the shapes of these spectra,
and $R_{xx}$ and $R_{x\delta}$ represent the effective size of ionized bubbles.  
In our analysis,
we adopt the values listed in Table~\ref{tab:Pxx_xdelta}
as the fiducial values of these parameters.

\begin{table}[t]
  \centering 
  \begin{tabular}{ccccccccc}
\hline \hline
~~$z$~~ & ~~$\bar{x}_H$~~
& ~~$b_{xx}^2$~~ & ~~$R_{xx}$~~  & ~~$\alpha_{xx}$~~ & ~~$\gamma_{xx}$~~ 
& ~~$b_{x\delta}^2$~~ & ~~$R_{x\delta}$~~  & ~~$\alpha_{x\delta}$~~ \\ 
&  &  & $[{\rm Mpc}]$& & & &$[{\rm Mpc}]$ \\
\hline
$9.2$  &  $0.9$ & $0.208$ & $1.24$ & $-1.63$ & $0.38$ & $0.45$ & $0.56$   & $-0.4$ \\ 
$8.0$  &  $0.7$ & $2.12$   & $1.63$ & $-0.1$   & $1.35$ & $1.47$ & $0.62$   & $0.46$ \\ 
$7.5$  &  $0.5$ & $9.9$     & $1.3$   & $1.6$    & $2.3$   & $3.1$   & $0.58$   & $2.0$ \\ 
$7.0$  &  $0.3$ & $77.0$   & $3.0$   & $4.5$    & $2.05$ & $8.2$   & $0.143$ & $28.0$ \\
\hline \hline
\end{tabular}
\caption{Fiducial values of the parameters characterizing $\mathcal{P}_{xx}(k)$
and $\mathcal{P}_{x\delta}(k)$ 
(See Eqs.~\eqref{eq:Pxx} and \eqref{eq:Pxdelta}) \cite{Mao:2008ug}.
}\label{tab:Pxx_xdelta}
\end{table}

%
We note that
the power spectrum in the $k$-space $P_{21} (\bm{k})$
are not directly measured by 21 cm line observations.
Instead, the angular location on the sky and the frequency 
are measured by an experiment,
and they can be specified by the following vector
\begin{align}
\bm{\Theta} = \theta_x \hat{e}_x + \theta_y \hat{e}_y + \Delta f \hat{e}_z 
\equiv  \bm{\Theta}_\perp + \Delta f \hat{e}_z,
\end{align}
where $\Delta f$ represents the frequency difference 
from the central redshift $z$ of a given redshift bin.
Then, we can define the Fourier dual of $\bm{\Theta}$ as
\begin{align}
{\bm u} \equiv u_x \hat{e}_x + u_y \hat{e}_y + u_\parallel \hat{e}_z 
\equiv {\bm u}_\perp + u_\parallel \hat{e}_z.
\end{align}
Notice that $u_\parallel$ has the unit of time
since it is the Fourier dual of $\Delta f$.
With the flat-sky approximation\footnote{
  Even if we consider all-sky experiments, the flat-sky approximation
  can be valid as long as we analyze 
  the data in a lot of small patches of the sky \cite{Mao:2008ug}.
}, 
we can linearize the relations between ${\bm r}$ and ${\bm \Theta}$,
and they are given by
\begin{align}
\bm{\Theta}_\perp = \bm{r}_\perp / d_A (z), 
\qquad
\Delta f = \Delta r_\parallel / y(z),
\end{align}
where $\bm{r}_\perp$ is the vector perpendicular to the line of sight,
$\Delta r_\parallel$ is the comoving distance interval 
corresponding to the frequency intervals $\Delta f$,
$d_A (z)$ is the comoving angular diameter distance, 
and $y(z) \equiv \lambda_{21} (1+z)^2 / H(z)$. 
Then, the relations between ${\bm k}$ and ${\bm u}$ 
can be written as
\begin{align}
\bm{u}_\perp = d_A \bm{k}_\perp, 
\qquad
u_\parallel = y k_\parallel.
\end{align}
The power spectrum of $\Delta T_b$ in the $u$-space can be defined 
in the same manner as the treatment in the $k$-space,
and the spectra are related each other by
\begin{align}
P_{21} (\bm{u})  = \frac{1}{d_A(z)^2  y (z)} P_{21} (\bm{k}).
\end{align}
We use the $u$-space power spectrum in the following analysis
because this quantity is directly measurable 
without assuming cosmological parameters.

\subsubsection{Fisher matrix of 21 cm line observation}\label{subsubsec:Forecast}

In order to estimate errors of cosmological parameters, 
we use the Fisher matrix analysis \cite{Tegmark:1996bz}.
The Fisher matrix of 21 cm line observations 
is given by \cite{McQuinn:2005hk}
\begin{align}
F^{({\rm 21cm})}_{\alpha\beta} = 
\sum_{\rm pixels}  
\frac{1}{[ \delta P_{21}(\bm{u}) ]^2} 
\left( \frac{\partial P_{21} (\bm{u})}{\partial \theta_{\alpha}} \right)
\left( \frac{\partial P_{21} (\bm{u})}{\partial \theta_{\beta}} \right),
\label{eq:Fisher_21}
\end{align}
where $\delta P_{21}(\bm{u})$ is the error 
in the power spectrum measurements for a Fourier pixel $\bm{u}$, 
and $\theta_{\alpha}$ represents a cosmological parameter with its index "$\alpha$".  
The 1 $\sigma$ error of the parameter $\theta_{\alpha}$ 
is evaluated by the Fisher matrix,
and we can obtain the estimated error from
\begin{align}
    \Delta \theta_{\alpha} \geq \sqrt{(F^{-1})_{\alpha\alpha}}.
\end{align}
When we differentiate $P_{21}({\bm u})$ with respect to 
the cosmological parameters, 
we fix ${\cal P}_{\delta \delta}(k)$ 
in Eqs.~(\ref{eq:Pxx})~and~(\ref{eq:Pxdelta})
so that we get conservative evaluations for 
errors of cosmological parameters.
Then,
the information of the matter distribution only 
comes from the ${\cal P}_{\delta \delta}(k)$ 
terms in $P_{\mu^0},P_{\mu^2}$ and $P_{\mu^4}$.

The error of the power spectrum
$\delta P_{21}({\bm u})$ consists of sample variances and experimental noises
%
\begin{align}
\delta P_{21}(\bm{u}) = \frac{ P_{21}(\bm{u})  + P_N (u_{\perp}) }{N_c^{1/2} },
\label{eq:variance_21cm}
\end{align}
where the first term on the right hand side 
represents the sample variance,
and the second term gives contribution of experimental noises. 
Here, $N_c = 2 \pi k_\perp \Delta k_\perp \Delta k_\parallel V(z) /(2\pi)^3$
is the number of independent cells in an annulus summing over the azimuthal angle,
$V(z) = d_A(z)^2 y(z) B \times {\rm FoV} $ is the survey volume,
where $B$ is the bandwidth,
and 
FoV $\propto \lambda^{2}$ is the field of view of an interferometer.
%

\subsubsection{Specifications of experiments}\label{sec:21cm_spec}

\begin{table}
\centering 
\begin{tabular}{c|ccccccc}
\hline \hline
& $N_{\rm ant}$ & $A_e~(z=8)$   & $L_{\rm min} $ & $ L_{\rm max}$ & FoV $(z=8)$     & Obs. time $t_0$ & z 
\\
&               & $[{\rm m}^2]$ & $[{\rm m}]$    & $[{\rm km}]$   & $[{\rm deg}^2]$ & [hour]          &  
\\ 
\hline
SKA phase 1 & 911/2         & 443 & 35 & 6 & 13.12                       & 1000 & 6.8 -- 10
\\
SKA phase 2 & 911$\times$ 4 & 443 & 35 & 6 & 13.12 $\times$ 4 $\times$ 4 & 1000 & 6.8 -- 10   
\\
Omniscope   & $10^6$        & 1   &  1 & 1 & 2.063 $\times 10^4$         & 1000 & 6.8 -- 10  
\\ 
\hline
\hline
\end{tabular}
\caption{Specifications of 21 cm observations adopted in the analysis. }
\label{tab:21obs}
\end{table}

Here, we show the specifications 
of the 21 cm line observations which is considered in this paper.
%
\paragraph{Survey range:}
%
In our analyses, we assume that the redshift range
used for constraining cosmological parameters
is $ z = 6.75 - 10.05$,
which we divide into 4 bins: $z = 6.75 - 7.25, 7.25 - 7.75, 7.75-8.25$
and $8.25 -  10.05$.  
For surveyed scales (wave number), 
we set a minimum cut off $k_{{\rm min} \parallel} = 2 \pi / (y B) $ 
in order to avoid foreground contaminations \cite{McQuinn:2005hk},
and take a maximum value $ k_{\rm max} = 2~{\rm Mpc}^{-1}$ 
in order not to be affected by nonlinear effects
of matter fluctuations,
which becomes important on $k_{\rm max} \le k$. 

For methods of foreground removals, see also recent discussions about
the independent component analysis (ICA) algorithm
(FastICA)~\cite{Chapman:2012yj}, which will be developed in terms of the
ongoing LOFAR observation~\cite{LOFAR}.

\paragraph{Noise power spectrum:}

The noise power spectrum, $P_N (u_{\perp})$ appeared
in Eq.~\eqref{eq:variance_21cm} is given by
\begin{align}
P_N (u_{\perp}) 
= \left( \frac{\lambda^{2} (z) T_{\rm sys} (z)  }{A_e (z)} \right)^2 
\frac{1}{t_0 n(u_\perp)},
\end{align}
where $\lambda$ is an observed wave length of redshifted 21 cm line,
$A_e$ is the effective collecting area per a antenna tile or a station,
$t_0$ is the observation time, 
$n(u_{\perp})$ is the number density of baseline,
and
the system temperature $T_{\rm sys}$ is estimated to be
$T_{\rm sys} = T_{{\rm sky}} + T_{{\rm rcvr}}$
and is dominated by the sky temperature due to synchrotron radiation.
Here, $T_{{\rm sky}} = 60 (\lambda/[m])^{2.55} $ [K] 
is the sky temperature, 
and $T_{{\rm rcvr}} = 0.1T_{{\rm sky}} + 40 $[K] 
is the receiver noise \cite{SKA}.
The  effective collecting area is proportional to 
the square of the observed wave length $A_e \propto \lambda^{2} $.
The number density of the baseline $n(u_\perp)$ 
depends on an antenna distribution.

As future observations of 21cm line fluctuations, in this paper
we consider SKA (phase~1 and phase~2) \cite{SKA,Mellema:2012ht}
and Omniscope \cite{Tegmark:2009kv,Tegmark:2008au},
whose specifications are shown in Table~\ref{tab:21obs}.
In order to estimate the number density of the baseline $n(u_\perp)$,
we assume a realization of antenna distributions for these arrays as follows.
The total collecting area of SKA phase~1 (SKA1)
is one-half as large as that of the originally planned SKA1.
Therefore, for SKA1,
we assume that the number of antenna station $N_{\textrm{ant}}$ is half as many as 
that of the originally planned SKA1, which has 911 antenna stations, 
and for SKA phase~2 (SKA2),  
the number of antenna stations is 4 times as many as that of the originally planned SKA1.

The number density of baselines of the originally planned SKA1
is determined as follows. 
%
We take the antenna stations 
in a core region with a radius 3000 m, 
which consists of 95\% of the total.
and the distribution has an antenna station density profile 
of the originally planned SKA1 $\rho_{{\rm origSKA1}}(r)$ 
($r$: a radius from center of the array) as follows \cite{Kohri:2013mxa},
\begin{align}
\rho_{{\rm origSKA1}}(r) = 
\left\{
\begin{array}{lll}
 \rho_{0}r^{-1},   &\rho_{0} \equiv \frac{13}{16\pi\left(\sqrt{10}-1\right) }  \ {\rm m}^{-2}
& \hspace{60pt} r \leq 400 \ {\rm m},\\
 \rho_{1}r^{-3/2}, &\rho_{1} \equiv \rho_{0} \times 400^{1/2}, & \ \ \ 400 \ {\rm m} < r \leq 1000 \ {\rm m}, \\
 \rho_{2}r^{-7/2}, &\rho_{2} \equiv \rho_{1} \times 1000^{2}, & \ \ 1000 \ {\rm m} < r \leq 1500 \ {\rm m}, \\
 \rho_{3}r^{-9/2}, &\rho_{3} \equiv \rho_{2} \times 1500 ,          & \ \ 1500 \ {\rm m} < r \leq 2000 \ {\rm m}, \\
 \rho_{4}r^{-17/2},&\rho_{4} \equiv \rho_{3} \times 2000^{4}, & \ \ 2000  \ {\rm m} < r \leq 3000 \ {\rm m}. \\
\end{array}
\right.
\end{align}
Here, we assume an azimuthally symmetric distribution of the antenna stations in SKA.
In this analysis, we ignore measurements from the sparse distribution of 
the remaining 5\% of the total antenna stations which are outside the core region.
This distribution agrees with the specification of 
the originally planned SKA1 baseline design.

Then we can evaluate the number density of baselines 
of the originally planned SKA1 $n_{{\rm origSKA1}}(u_\perp)$
from this distribution.
Using the number density of baselines,
we can estimate that the number densities of baselines 
of SKA1~(re-baseline design) and SKA2 are
\begin{align}
n_{{\rm SKA1}}(u_\perp) 
&= n_{{\rm origSKA1}}(u_\perp) \times \left( \frac{1}{2} \right)^{2}, \\
n_{{\rm SKA2}}(u_\perp)
&= n_{{\rm origSKA1}}(u_\perp) \times 4^{2},
\end{align}
where $n_{{\rm SKA1}}(u_\perp)$ 
and $n_{{\rm SKA2}}(u_\perp)$
are the number densities of baseline of SKA1
or SKA2, respectively.

For Omniscope, which is a future square-kilometre collecting area array
optimized for 21 cm tomography, we take all of antenna tiles distributed 
with a filled nucleus in the same manner as Ref. \cite{Mao:2008ug}.

\subsection{CMB}

In our analysis, 
we focus on not only 21cm line observations
but also CMB observations, 
especially gravitational lensing of CMB,
which has information on matter fluctuations at late times.
Although the 21 cm line observations are
powerful probes of the matter power spectrum, 
particularly, on small scales, 
the CMB observations greatly help to determine 
other cosmological parameters
such as energy densities of the dark matter, 
baryons and the dark energy.

Besides, CMB power spectra are sensitive to gravitino mass 
through the CMB lensing.
Future precise CMB experiments are expected to 
set stringent constraints on the light gravitino mass~\cite{Ichikawa:2009ir}.
Therefore, we take account of 
combining the CMB experiments 
with the 21 cm line observations.  

\subsubsection{Fisher matrix of CMB}

We evaluate errors of cosmological parameters
by using the Fisher matrix of CMB,
which is given by~\cite{Tegmark:1996bz}
\begin{align}
F_{\alpha \beta}^{\rm (CMB)}
= \sum_{\ell}
\frac{\left( 2\ell+1\right)}{2}
\mathrm{Tr}
\left[
  C_{\ell}^{-1}
  \frac{\partial C_{\ell}}{\partial \theta_{\alpha}}
  C_{\ell}^{-1}
  \frac{\partial C_{\ell}}{\partial \theta_{\beta}}
\right],
\label{eq:Fisher_CMB}
\end{align}
\begin{align}
    C_{\ell} =  \left(
\begin{array}{ccc}
\hspace{5pt} C_{\ell}^{\mathrm{TT}} 
+ N_{\ell}^{\mathrm{TT}} \hspace{5pt} &
\hspace{5pt} C_{\ell}^{\mathrm{TE}} \hspace{5pt} &
\hspace{5pt} C_{\ell}^{\mathrm{Td}} \hspace{5pt} \\
\hspace{5pt} C_{\ell}^{\mathrm{TE}} \hspace{5pt} &
\hspace{5pt} C_{\ell}^{\mathrm{EE}} 
+N_{\ell}^{\mathrm{EE}} \hspace{5pt} &
0 \hspace{5pt} \\
\hspace{5pt} C_{\ell}^{\mathrm{Td}} \hspace{5pt} &
0 \hspace{5pt} &
\hspace{5pt} C_{\ell}^{\mathrm{dd}} 
+ N_{\ell}^{\mathrm{dd}} \hspace{5pt}
\end{array}
\right),
\end{align}
where $\ell$ is the multipole of angular power spectra,
$C_{\ell}^{\mathrm{X}} 
\left(\mathrm{X}=\mathrm{TT, EE, TE} \right)$
are the CMB power spectra,
$C_{\ell}^{\mathrm{dd}}$ is the deflection angle spectrum,
$C_{\ell}^{\mathrm{Td}}$ is the
cross correlation between the deflection angle and the temperature,
$N_{\ell}^{\mathrm{X'}}$ 
$\left(\mathrm{X'}=\mathrm{TT, EE} \right)$
and $N_{\ell}^{\mathrm{dd}}$
are the noise power spectra, 
where $C_{\ell}^{\mathrm{dd}}$ is calculated by a lensing
potential~\cite{Okamoto:2003zw} and is related with
the lensed CMB power spectra.
The noise power spectra of CMB 
$N_{\ell}^{\mathrm{X'}}$ are expressed with a beam size
$\sigma_{\mathrm{beam}}(\nu)=$ $\theta
_{\mathrm{FWHM}}(\nu)/\sqrt{8\ln 2}$,
where 
$\sqrt{8\ln 2}\sigma_{{\rm beam}}$ means 
the full width at half maximum of the Gaussian distribution,
and instrumental sensitivity
$\Delta _{\mathrm{X'}}(\nu)$ by
\begin{align}
  N_{\ell}^{\mathrm{X'}}
  = \left[ 
    \sum_{i} \frac1{n_{\ell}^{\mathrm{X'}}(\nu_{i})}
    \right]^{-1},
    \label{eq:noise_CMB}
\end{align}
where $\nu_{i}$ is an observing frequency and
\begin{align}
n_{\ell}^{\mathrm{X'}}(\nu)=
\Delta ^2_{\mathrm{X'}}(\nu)\exp\left[\ell(\ell+1) \sigma ^2_{\mathrm{beam}}(\nu)\right].
\end{align}
The noise power spectrum of deflection angle 
$N^{dd}_l$ is obtained assuming lensing  reconstruction with the
quadratic estimator \cite{Okamoto:2003zw},
which is computed with FUTURCMB \cite{paper:FUTURCMB}.
In this algorithm,
$N_{\ell}^{dd}$ is estimated from
the noise $N_{\ell}^{X'}$, 
and 
lensed and unlensed power spectra of CMB temperature, 
E-mode and B-mode polarizations.

Finally, the Fisher matrix in Eq.\eqref{eq:Fisher_CMB} is modified as follows 
by taking into account the multipole range 
[$\ell_{min}$, $\ell_{max}$] 
and the fraction of the observed sky $f_{{\rm sky}}$,
\begin{equation}
F_{\alpha \beta}^{\rm (CMB)}
= \sum_{\ell = \ell_{min}}^{\ell_{max}}
\frac{\left( 2\ell+1\right)}{2}f_{\mathrm{sky}}
\mathrm{Tr}
\left[
  C_{\ell}^{-1}
  \frac{\partial C_{\ell}}{\partial \theta_{\alpha}}
  C_{\ell}^{-1}
  \frac{\partial C_{\ell}}{\partial \theta_{\beta}}
\right].
\end{equation}

\subsubsection{Specifications of the experiments}

Now, we show the specifications of 
the CMB observations
which are considered in this paper.
In order to obtain the future constraints, 
we consider Planck~\cite{Planck:2006aa},
the Simons Array\cite{SimonsArray}, which we occasionally abbreviate to SA in this paper,
and COrE+ \cite{COrE+}
whose experimental specifications 
are summarized in Tables~\ref{tab:Planck_SA_spec} and \ref{tab:core_spec}.
The Simons Array is a near future
ground-based precise CMB polarization observation
and COrE+ is a planned satellite observing CMB.

%
%
%
%

When we combine observations of Planck and the Simons Array, we
evaluate the noise power spectra 
$N_{\ell}^{\mathrm{X,Planck+SA}}$ of 
the CMB polarization ($X = {\rm EE}$ or ${\rm BB}$)
with the following operation:
\begin{description}
\item[(1)] $2\leq \ell < 25$ 
\begin{align}
N_{\ell}^{\mathrm{X,Planck + SA}} 
= N_{\ell}^{\mathrm{X,Planck}},
\end{align}
\item[(2)] $25\leq \ell \leq 3000$
\begin{align}
N_{\ell}^{\mathrm{X,Planck + SA}} 
= [1/N_{\ell}^{\mathrm{X,Planck}} 
+ 1/N_{\ell}^{\mathrm{X,SA}}]^{-1}.
\end{align}
\end{description}
Since we assume that 
the CMB temperature fluctuation observed by the Simons Array 
is not used for constraints on the cosmological parameters,
the temperature noise power spectrum 
$N_{\ell}^{\mathrm{TT,\,Planck+SA}}$ is equal to 
{\bf $N_{\ell}^{\mathrm{TT,\,Planck}}$}.
%
The reason for this
is that the CMB temperature fluctuation 
observed by Planck reaches almost cosmic variance limit.
%
Therefore, the constraints are not significantly improved
if we include the CMB temperature fluctuation
observed by  the Simons Array.

\begin{table}[tp]
\begin{center}
\begin{tabular}{c|ccccccc}
\hline
\hline
\shortstack{Experiment\\ \,}&
\shortstack{$\nu$ \\ $[\mathrm{GHz}]$}&
\shortstack{$\Delta _{\mathrm{TT}}$\\ $[\mathrm{\mu K\textrm{arcmin}}]$}&
\shortstack{$\Delta _{\mathrm{PP}}$\\ $[\mathrm{\mu K\textrm{arcmin}}]$}& 
\shortstack{$\theta_{\mathrm{FWHM}}$\\ $[\mathrm{\textrm{arcmin}}]$}&
\shortstack{$f_{\mathrm{sky}}$\\ $ $} &
\shortstack{$\ell_{{\rm min}}$\\ $ $} &
\shortstack{$\ell_{{\rm max}}$\\ $ $} 
\\
\hline
\hline
       & 30 & 145 & 205  & 33  &      &   &      \\ 
       & 44 & 150 & 212  & 23  &      &   &      \\ 
       & 70 & 137 & 195  & 14  &      &   &      \\
Planck & 100& 64.6& 104  & 9.5 & 0.65 & 2 & 3000 \\
       & 143& 42.6& 80.9 & 7.1 &      &   &      \\
       & 217& 65.5& 134  & 5   &      &   &      \\ 
       & 353& 406 & 406  & 5   &      &   &      \\ 
\hline
\hline
              & 95  & - & 13.9 & 5.2 &       &    &      \\ 
Simons Array  & 150 & - & 11.4 & 3.5 & 0.65  & 25 & 3000 \\
              & 220 & - & 38.8 & 2.7 &       &    &      \\                 
\hline
\hline
\end{tabular}
\caption{
Experimental specifications of Planck and the Simons Array 
assumed in our analysis~\cite{Oyama:2015gma}. 
Here $\nu$ is the observation frequency, 
$\Delta_{\rm TT}$ is the temperature sensitivity per $1'\times1'$
pixel, $\Delta_{\rm PP}=\Delta_{\rm EE}=\Delta_{\rm BB}$ 
is the polarization (E-mode and B-mode) sensitivity 
per $1'\times1'$ pixel,
$\theta_{\rm FWHM}$ is the angular resolution defined to be the full width at
half-maximum, and $f_{\rm sky}$ is the observed fraction of the sky.
For the Planck experiment, we assume that 
the three frequency bands ($70, 100, 143$ GHz)
are only used for the observation of CMB.
For the Simons Array, we assume that
the 95 and 150 GHz bands are used for observation of CMB,
and the 220 GHz band is not used for constraining cosmological parameters
because the band is found to be useful for the foreground removal~\cite{Oyama:2015vva,Oyama:2015gma}.
}
\label{tab:Planck_SA_spec}
\end{center}
\end{table}

\begin{table}
\begin{tabular}{c|ccccccc}
\hline
\hline
\shortstack{Experiment\\ \,}&
\shortstack{$\nu$ \\ $[\mathrm{GHz}]$}&
\shortstack{$\Delta _{\mathrm{TT}}$\\ $[\mathrm{\mu K\textrm{arcmin}}]$}&
\shortstack{$\Delta _{\mathrm{PP}}$\\ $[\mathrm{\mu K\textrm{arcmin}}]$}& 
\shortstack{$\theta_{\mathrm{FWHM}}$\\ $[\mathrm{\textrm{arcmin}}]$}&
\shortstack{$f_{\mathrm{sky}}$\\ $ $} &
\shortstack{$\ell_{{\rm min}}$\\ $ $} &
\shortstack{$\ell_{{\rm max}}$\\ $ $} 
\\
\hline
\hline
     &45   & 5.2 & 9.0 & 23.3 & & & \\
     &75   & 2.7 & 4.7 & 14   & & &  \\
     &105  & 2.7 & 4.6 & 10   & & &  \\
     &135  & 2.6 & 4.5 & 7.7  & & &  \\
     &165  & 2.6 & 4.6 & 6.4  & & &  \\
     &195  & 2.6 & 4.5 & 5.4  & & &  \\
     &225  & 2.6 & 4.5 & 4.7  & & &  \\
COrE+&255  & 6.0 & 10.4& 4.1  & 0.65 & 2 & 3000  \\
     &285  & 10.0& 17  & 3.7  & & &  \\
     &315  & 26.6& 46  & 3.3  & & &  \\
     &375  & 67.8& 117 & 2.8  & & &   \\
     &435  & 147.6&255    & 2.4 & & & \\                
     &555  & 218  & 589   & 1.9 & & & \\                
     &675  & 1268 & 3420  & 1.6 & & & \\                
     &795  & 7744 & 20881 &1.3  & & & \\                  
\hline \hline
\end{tabular}
  \centering 
  \caption{Experimental specifications of COrE+ adopted in our analysis.
Because specifications of COrE+ are in the planning stage,
we use the values appeared in Ref.~\cite{COrE+},
which are original specifications of COrE.
In the same manner as Ref.~\cite{COrE+},
we assume that the CMB channels are 75, 105, 135, 165, 195 and 225~GHz.}
\label{tab:core_spec}
\end{table}

\subsection{BAO}

In our analysis, we take account of joint constraints from
CMB, 21cm line, baryon acoustic oscillation (BAO). 
Therefore, before discussing future constraints,
we briefly summarize formalisms of our analysis methods related to 
BAO, here. 
%

\subsubsection[Fisher matrix of BAO]{Fisher matrix of BAO }
\label{subsubsec:BAOFisher}

For estimating future constrains,
we use the following Fisher matrix of BAO data.
The following method is based on 
\cite{Albrecht:2006um} and \cite{Wu:2014hta}. 
The observables of BAO are the comoving angular diameter
distance $d_{A}(z)$ and the Hubble parameter $H(z)$
(and more specifically, $\ln(d_{A}(z))$ and $\ln(H(z))$ are the observables).
Therefore, the Fisher matrix of BAO data is written as
\begin{align}
F^{(\textrm{BAO}) \ d,H}_{\alpha \beta} &=
\sum_{i} \frac{1}{\sigma_{d,H}^2(z_{i})} 
\frac{\partial f_{i}^{d,H}}{\partial \theta_{\alpha}}
\frac{\partial f_{i}^{d,H}}{\partial \theta_{\beta}}, \\
f_{i}^{d} &= \ln(d_{A}(z_{i})), \\
f_{i}^{H} &= \ln(H(z_{i})),
\end{align}
where $i$ is the index of each redshift bin,
$\sigma_{d}(z_{i})$ and $\sigma_{H}(z_{i})$
are variances of $\ln(d_{A}(z_{i}))$
and $\ln(H(z_{i}))$, respectively. 
We assume that an observed redshift range is divided into bins,
with the width and central redshift value of each bin respectively denoted as $\Delta z_i$ and $z_i$.

Note that  cosmological parameters related to BAO data
are only $(\Omega_{m}h^2, \Omega_{\Lambda})$
or $(h, \Omega_{\Lambda})$ 
when we assume that the Universe is flat.

\subsubsection{Specifications of BAO data and the direct measurement of the Hubble constant}

\begin{table}
\begin{tabular}{c|c|c}
\hline \hline 
 Central redshift $z_{i}$ & $\sigma_{d}(z_{i}) \times 10^{2}$  &  $\sigma_{H}(z_{i})\times 10^{2}$ \\ 
\hline
0.15 & 2.78 & 5.34 \\
0.25 & 1.87 & 3.51 \\
0.35 & 1.45 & 2.69 \\
0.45 & 1.19 & 2.20 \\
0.55 & 1.01 & 1.85 \\
0.65 & 0.87 & 1.60 \\
0.75 & 0.77 & 1.41 \\
0.85 & 0.76 & 1.35 \\
0.95 & 0.88 & 1.42 \\
1.05 & 0.91 & 1.41 \\
1.15 & 0.91 & 1.38 \\
1.25 & 0.91 & 1.36 \\
1.35 & 1.00 & 1.46 \\
1.45 & 1.17 & 1.66 \\
1.55 & 1.50 & 2.04 \\
1.65 & 2.36 & 3.15 \\
1.75 & 3.62 & 4.87 \\
1.85 & 4.79 & 6.55 \\
\hline \hline
\end{tabular}
  \centering 
  \caption{Specification of DESI (14000 [${\rm deg}^2$]) adopted in the analysis.
These values are the same as those in Ref.\cite{Font-Ribera:2013rwa}. }
  \label{tab:DESI_spec}
\end{table}

In this paper, we focus on the 
Dark Energy Spectroscopic Instrument (DESI) \cite{DESI:web,Font-Ribera:2013rwa},
which is a future large volume galaxy survey. 
The survey redshift range is $0.1<z<1.9$ 
(we do not include the Ly-$\alpha$ forest at $1.9<z$ for simplicity),
where we assume that the redshift range is divided into 18 bins, 
in other words $\Delta z_{i}=0.1$,
and the observed solid angle is
$14000$~$[{\rm deg}^2]$.
We summarize the specifications of DESI in
Table.~\ref{tab:DESI_spec} \cite{Font-Ribera:2013rwa}.

Additionally, in the same manner as \cite{Wu:2014hta},
when we combine BAO with the other observations,
we add a 1\% $H_{0}$ prior,
which would be achievable by a direct measurements of the Hubble constant
in the next decade.
The Fisher matrix of the direct measurement of $H_0$ is expressed as 
\begin{align}
F^{(H_{0})}_{\alpha \beta} 
= \left\{
\begin{array}{c}
 \hspace{-40pt} \frac{1}{(1\%\times H_{0,{\rm fid}})^2}, 
 \hspace{30pt} \theta_{\alpha}=\theta_{\beta}=H_{0},\\
 \hspace{20pt}0,
 \hspace{60pt} {\rm the \ other \ components},
\end{array}
\right.
\end{align}
where $H_{0,{\rm fid}}$ is the fiducial value of $H_{0}$.
If we choose the Hubble parameter as 
a dependent parameter,
it is necessary to translate the Fisher matrix
into that of a chosen parameter space.
Under the translation of 
$(h, \Omega_{\Lambda})$ $\longrightarrow$
$(\Omega_{m}h^2, \Omega_{\Lambda})$,
the Fisher matrix in the new parameter space is written as
\begin{align}
\tilde{F}^{H_{0}} = 
\left(
\begin{array}{cc}
\tilde{F}_{\Omega_{m}h^2 \Omega_{m}h^2} &
\tilde{F}_{\Omega_{m}h^2 \Omega_{\Lambda}} \\
\tilde{F}_{\Omega_{m}h^2 \Omega_{\Lambda}} & 
\tilde{F}_{\Omega_{\Lambda} \Omega_{\Lambda}}
\end{array}
\right)
=
\frac{1}{(1\%\times H_{0,{\rm fid}})^2}
\left(\frac{1}{2\Omega_{m}h^2}\right)^2
\left(
\begin{array}{cc}
h^2 & h^4 \\
h^4 & h^6
\end{array}
\right).
\end{align}
%

\section{Results}
\label{sec:results}

In this section, we present our results for 
projected constraints 
by 21cm line~(SKA phase~1, phase~2, or Omniscope), 
CMB~(Planck + Simons Array (SA) or COrE+), 
BAO~(DESI) and a direct measurement of the Hubble constant
on cosmological parameters, 
paying particular attention to 
parameters related to the light gravitino, 
i.e. 
the fraction of light gravitinos 
in the total dark matter density $f_{3/2}$,
and the effective number of neutrino species 
for light gravitinos $N_{3/2}$.

When we calculate the Fisher matrices, we choose the following basic set
of cosmological parameters:
the energy density of matter $\Omega_{m}h^{2}$, baryons $\Omega_{b}h^{2}$
and the dark energy $\Omega_{\Lambda}$, 
the scalar spectral index $n_{s}$, the scalar fluctuation amplitude $A_{s}$ 
(the pivot scale is taken to be $k_{{\rm pivot}}=$ $0.05 \ {\rm Mpc}^{-1}$), 
the reionization optical depth $\tau$, 
and the primordial value of the $^4$He mass fraction $Y_{\textrm{p}}$.
Fiducial values of these parameters 
are taken to
$(\Omega_{m}h^{2},\Omega_{b}h^{2},\Omega_{\Lambda},n_{s},A_{s}\times10^{10},\tau,Y_{\textrm{p}})$
$=( 0.1417, 0.0223, 0.6911, 0.9667, 21.42
, 0.066, 0.25)$,
which are the best fit values of the Planck result \cite{Ade:2015xua}.
For the total neutrino mass $\Sigma m_{\nu} = m_{1}+m_{2}+m_{3}$,
we fix $\Sigma m_{\nu}$ to a fiducial value $\Sigma m_{\nu} = 0.06 \ \textrm{eV}$,
or vary it freely.
%
In the following analysis, 
we fix the neutrino mass hierarchy to be the normal one
and the effective number of neutrino species $N_{\nu}$ to be $3.046$.
For the parameters related to the light gravitino,
we set the fiducial value of $N_{3/2}$ to be $0.059$
and that of $f_{3/2}$ to be
$0.01071$ or $0.05353$, which corresponds to $m_{3/2}=1$~eV
or $5$~eV, respectively when we fix $N_{3/2}$ to be $0.059$.
For $N_{3/2}$,
we fix $N_{3/2}=0.059$, or treat it as a free parameter.

To obtain Fisher matrices we use CAMB \cite{Lewis:1999bs,CAMB}\footnote{
In our analysis of 21 cm line, 
we neglect non-linear effects for evolutions of the matter power spectrum
because  
we adopt a 21 cm power spectrum only at linear regime.
For CMB lensing,
by performing a public code HALOFIT~\cite{Lewis:1999bs,CAMB},
we have checked that modifications
by including the non-linear effects
are much smaller than typical errors in our analyses 
and have negligible impacts on our constraints.
}
for calculations of CMB anisotropies $C_{l}$ and matter power spectra $P_{\delta \delta}(k)$.
In order to combine the CMB experiments with the 21 cm line, BAO
and a Hubble $H_0$ measurement,
we calculate the combined Fisher matrix 
\begin{equation}
F_{\alpha\beta} 
= F^{\rm (21cm)}_{\alpha\beta} + F^{\rm (CMB)}_{\alpha\beta} 
+ F^{\rm (BAO)}_{\alpha\beta} + \tilde{F}^{H_0}_{\alpha \beta}.
\end{equation}
In this paper, we do not use information for a possible correlation 
between fluctuations of the 21 cm and the CMB.

\subsection{ Expected future constraints on light gravitino}
\label{subsec:constraints}

In Tables.~\ref{tab:gravi1eV_fixN32}-\ref{tab:gravi5eV_mnu_free},
we summarize constraints on cosmological parameters
from each combination of experiments.

In Tables~\ref{tab:gravi1eV_fixN32} and \ref{tab:gravi5eV_fixN32},
the constraints are 
for the cases with fixed $N_{3/2}=0.059$ and $\Sigma m_{\nu}=0.06$~eV.
With Planck, the Simons Array, DESI~(BAO) and a direct measurement of $H_0$ combined,
we obtain a 1~$\sigma$ error on $f_{3/2}$ 
\begin{align}
\sigma(f_{3/2}) =& 0.00346,  
\end{align}
for fiducial $f_{3/2}=0.01071$, which corresponds to $m_{3/2}=1$~eV.
Adding 21 cm line experiments to them,
we see that the constraint on $f_{3/2}$ can be significantly improved. 
For example,
for fiducial $f_{3/2}=0.01071$, the combination of 
SKA phase~1 and Planck + Simons Array + DESI + $H_0$ gives
%
\begin{align}
\sigma(f_{3/2}) =& 0.00263,  
\end{align}
while the ones of SKA phase~2 and Planck + Simons Array + DESI + $H_0$
gives
%
%
\begin{align}
\sigma(f_{3/2}) =& 0.00165.  
\end{align}

We can translate them into errors on the mass of light gravitino $m_{3/2}$.
For the case with Planck, the Simons Array, DESI and $H_0$ combined,
the error of $m_{3/2}$ is given as
\begin{align}
\sigma(m_{3/2}) =& 0.33~\textrm{eV}. 
\end{align}
If we add SKA~phase~1 or phase~2 to them, 
the error can be improved as
\begin{align}
\sigma(m_{3/2}) =& 0.25~\textrm{eV} \ \ \ \textrm{(SKA phase~1)},  
\end{align}
\begin{align}
\sigma(m_{3/2}) =& 0.16~\textrm{eV} \ \ \ \textrm{(SKA phase~2)}.  
\end{align}
If we 
combine Omniscope with Planck + Simons Array + DESI + $H_0$, 
the error can improved even further as
\begin{align}
\sigma(m_{3/2}) =& 0.067~\textrm{eV}  \ \ \ \textrm{(Omniscope)}. 
\end{align}
Thus, from these strong improvements, 
we find that observations of 21 cm line are significantly 
useful to constrain the mass of light gravitinos.

\begin{figure}[htbp]
  \centering
  \includegraphics[bb=47 142 566 654,width=1\linewidth]{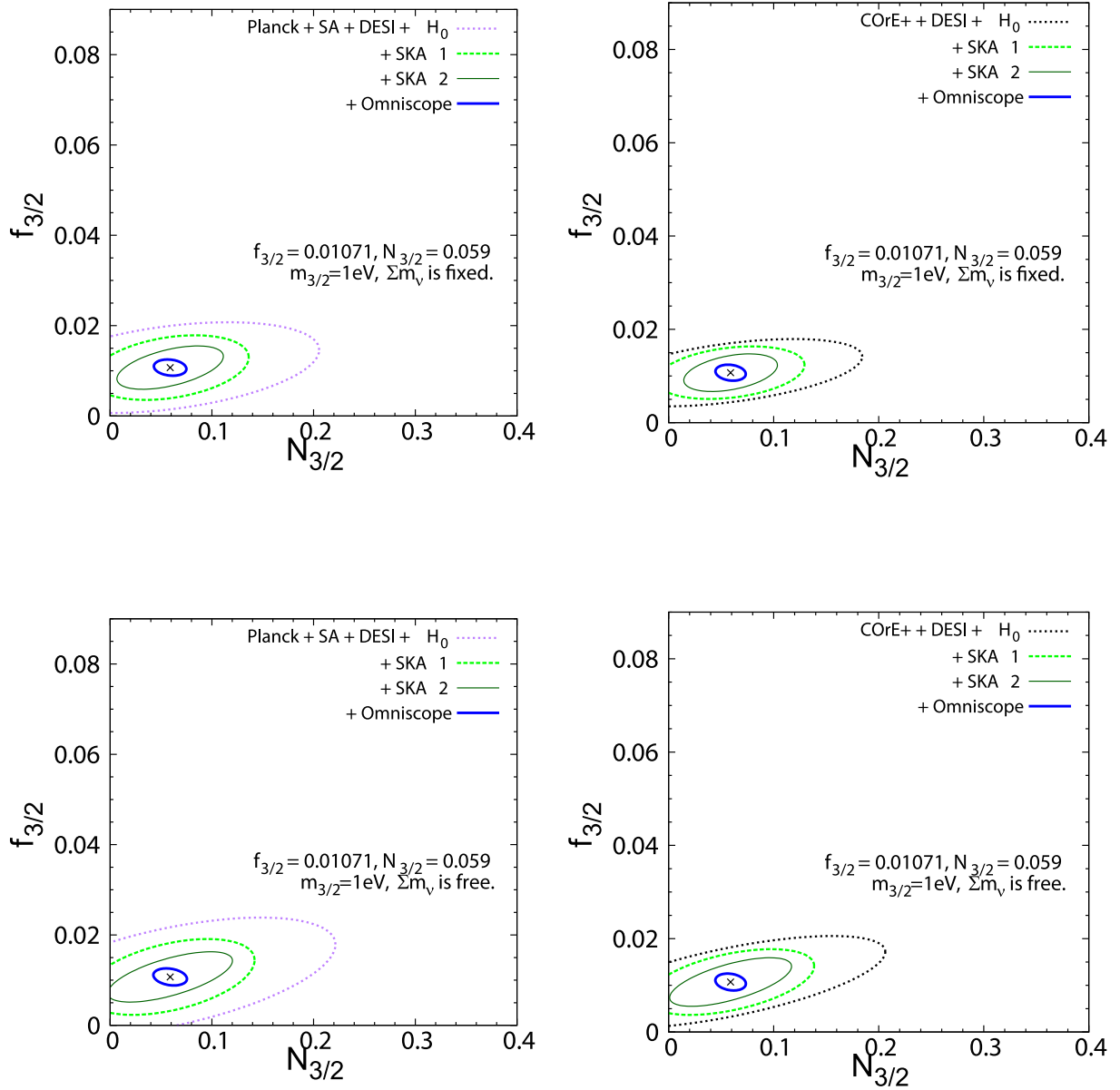}
  \caption{
Contours of 95\% C.L. forecasts in $N_{3/2}$-$f_{3/2}$ plane.
We assume that $f_{3/2}=0.01071$ and $N_{3/2}=0.059$, which correspond to $m_{3/2}=1$~eV.
In the upper panels, we fix the total neutrino mass $\Sigma m_{\nu}$, 
and in the lower panels, we treat the total neutrino mass as a free parameter.
We show constraints from Planck + Simons Array~(SA) + DESI~(BAO) + $H_{0}$ (dotted purple line)
with SKA phase 1 (dashed yellow-green line), phase 2 (solid green line) 
or Omniscope (thick blue line) in the left panels, 
and COrE+ + DESI~(BAO) + $H_{0}$ 
(dotted black line) with SKA phase 1, phase 2 or Omniscope in the right panels.
}\label{fig:gravitino_1eV}
\end{figure}

\begin{figure}[htbp]
  \centering
  \includegraphics[bb=47 142 566 654,width=1\linewidth]{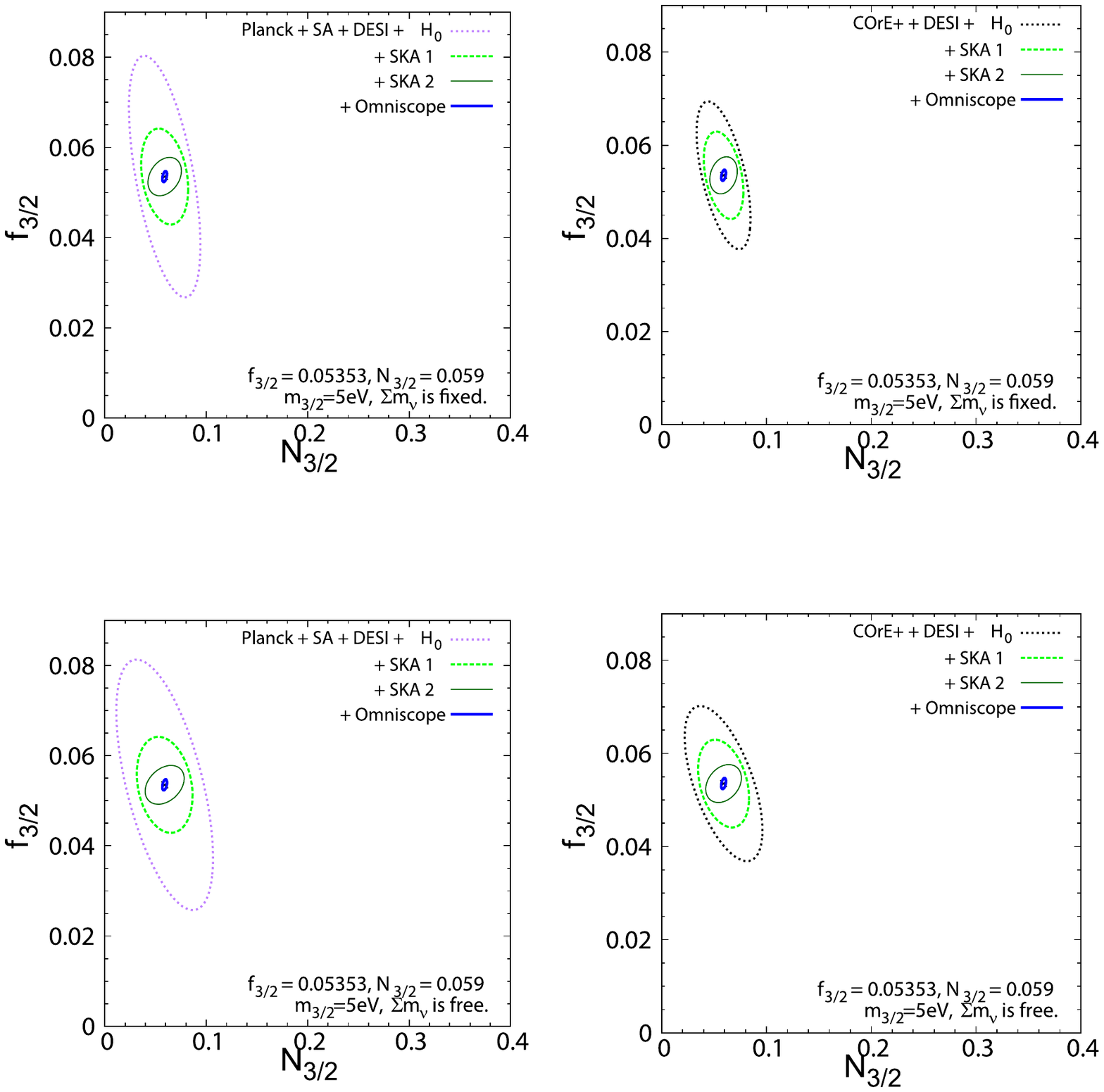}
  \caption{
The same as Fig.\ref{fig:gravitino_1eV} but for $f_{3/2}=0.05353$ 
and $N_{3/2}=0.059$, which correspond to $m_{3/2}=5$~eV.
}\label{fig:gravitino_5eV}
\end{figure}

\begin{figure}[htbp]
  \centering
  \includegraphics[bb=47 142 566 654,width=1\linewidth]{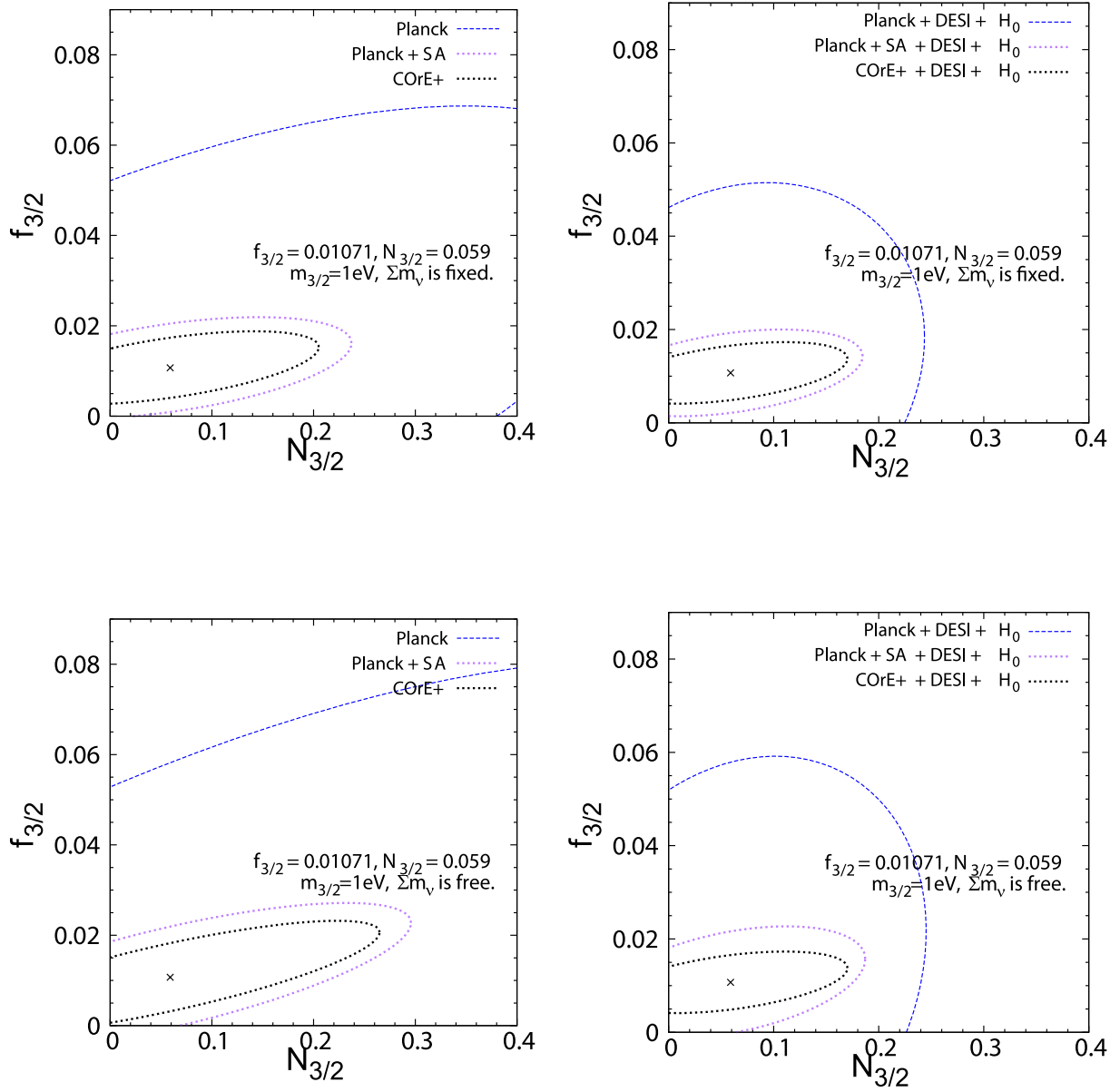}
  \caption{
Contours of 95\% C.L. forecasts in $N_{3/2}$-$f_{3/2}$ plane
by CMB combined with BAO~(DESI) and $H_{0}$.
We assume that $f_{3/2}=0.01071$ and $N_{3/2}=0.059$, which correspond to $m_{3/2}=1$~eV.
In the upper panels, we fix the total neutrino mass $\Sigma m_{\nu}$, 
and in the lower panels, we treat the total neutrino mass as a free parameter.
We show constraints from CMB only (Planck, Planck + Simons Array(SA) and COrE+) in the left panels, 
and combinations of CMB, BAO~(DESI) and $H_{0}$ in the right panels.
}\label{fig:gravitino_1eV_CMB}
\end{figure}

\begin{figure}[htbp]
  \centering
  \includegraphics[bb=47 142 566 654,width=1\linewidth]{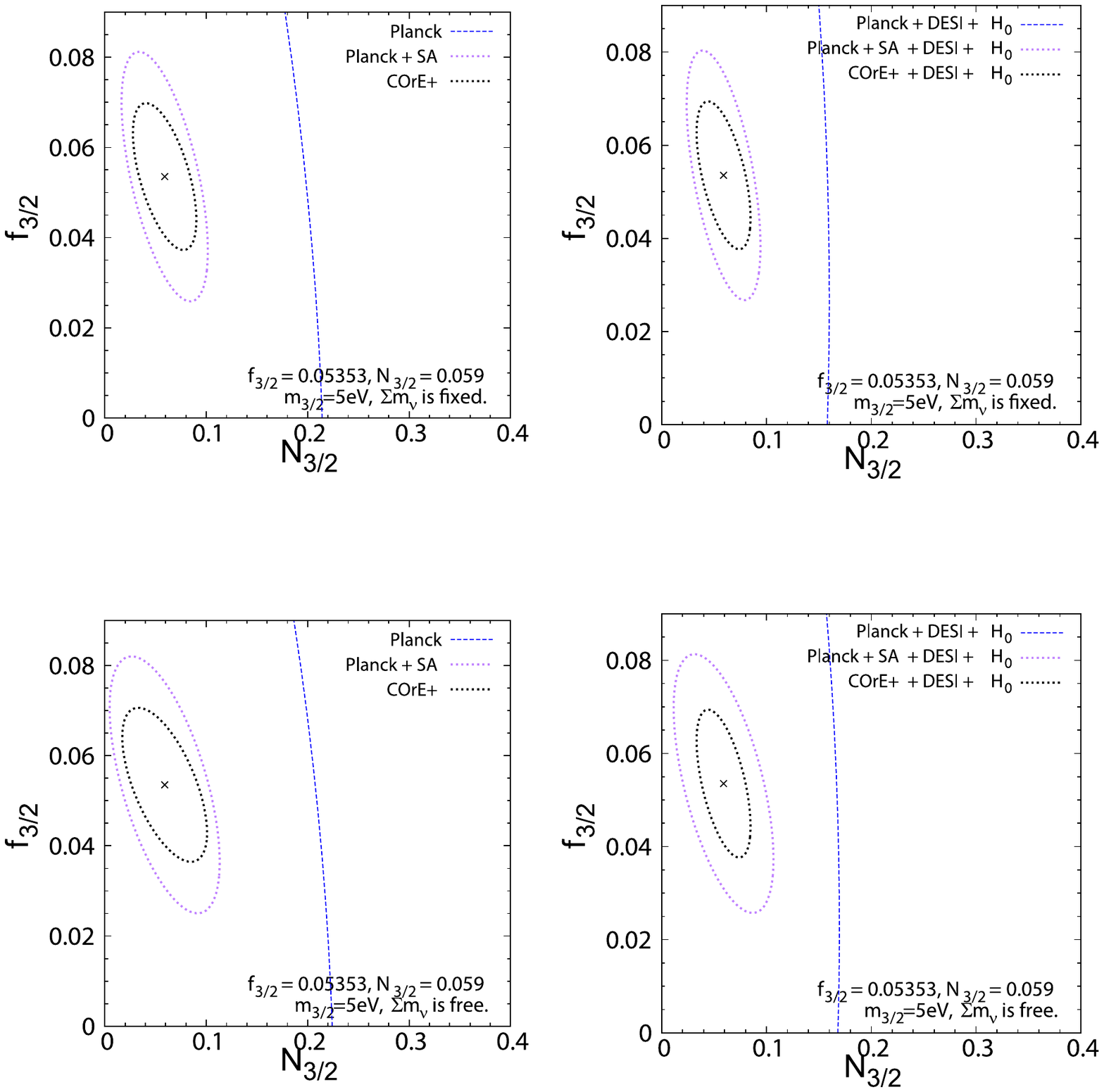}
  \caption{
The same as Fig.\ref{fig:gravitino_1eV_CMB} but for $f_{3/2}=0.05353$ 
and $N_{3/2}=0.059$, which correspond to $m_{3/2}=5$~eV.
}\label{fig:gravitino_5eV_CMB}
\end{figure}

Next, in Figs.~\ref{fig:gravitino_1eV}-\ref{fig:gravitino_5eV_CMB},
we plot contours of 95\% confidence levels (C.L.) forecasts 
of each combination of CMB, 21cm line, BAO and $H_0$
in $N_{3/2}$-$f_{3/2}$ plane. 
%
%
In 
the upper panels of Figs.~\ref{fig:gravitino_1eV}-\ref{fig:gravitino_5eV_CMB},
we fix the total neutrino mass $\Sigma m_{\nu}$,
and in those of the lower panels,
we treat $\Sigma m_{\nu}$ as a free parameter.
In Figs.~\ref{fig:gravitino_1eV} and \ref{fig:gravitino_1eV_CMB}
the fiducial value of $f_{3/2}$
is set to $f_{3/2} = 0.01071$, 
which corresponds to $m_{3/2}=1$~eV when we fix $N_{3/2}$ to be $0.059$.
In Figs.~\ref{fig:gravitino_5eV} and \ref{fig:gravitino_5eV_CMB},
the fiducial value of $f_{3/2}$
is set to $f_{3/2} = 0.05353$,
which corresponds to $m_{3/2}=5$~eV when we fix $N_{3/2}$ to be $0.059$.
In the left two panels of Figs.~\ref{fig:gravitino_1eV}-\ref{fig:gravitino_5eV},
each contour represents constraints from Planck + Simons Array + BAO~(DESI) + $H_0$ measurement
or Planck + Simons Array + BAO~(DESI) + $H_0$ measurement + 21cm line (SKA phase~1, phase~2 or Omniscope).
In the right two panels of them, each contour represents constraints
from COrE+ + BAO~(DESI) + $H_0$ measurement
or COrE+ + BAO~(DESI) + $H_0$ measurement + 21cm line (SKA phase~1, phase~2 or Omniscope).
In the left two panels of Figs.~\ref{fig:gravitino_1eV_CMB}-\ref{fig:gravitino_5eV_CMB},
each contour represents constraints by CMB (Planck, Planck + Simons Array or COrE+) only. 
In the right two panels of them, each contour represents constraints
by CMB (Planck, Planck + Simons Array or COrE+) + BAO~(DESI) + $H_0$ measurement.

From Figs.~\ref{fig:gravitino_1eV}-\ref{fig:gravitino_5eV_CMB},
we can see that constraints on $N_{3/2}$ and $f_{3/2}$
depend on the fiducial value of $f_{3/2}$.
As the fiducial value of $f_{3/2}$ becomes smaller,
the constraints on $f_{3/2}$  become better
while those
on $N_{3/2}$ become worse.
The dependences result from the following reasons.
From Eq.\eqref{eq:effective_num_gravi} and \eqref{eq:Omega_3_2},
the mass of light gravitinos behaves as
\begin{align}
m_{3/2} \propto
f_{3/2}N_{3/2}^{-\frac{3}{4}}. 
\end{align}
From this equation,
we can find that 
the variation of $m_{3/2}$ due to changing $N_{3/2}$
becomes smaller
as the fiducial value of $f_{3/2}$ becomes smaller.
Therefore, 
the influence due to changing $N_{3/2}$
on the growth of perturbations 
becomes less significant 
if the fiducial value of $f_{3/2}$ is small.
On the other hand, we can see that
the constraints on $f_{3/2}$ 
depend on the fiducial value of $f_{3/2}$
mainly in CMB observation. 
As the fiducial value of $f_{3/2}$ becomes larger,
the free streaming scale of light gravitinos becomes shorter,
which makes
it more difficult
to obtain the information of the gravitino mass from CMB observations
because we need to measure 
higher multi-pole power spectra $C_{\ell}$
in order to obtain the information
of the free-streaming scales.

Next, from Figs.~\ref{fig:gravitino_1eV_CMB}-\ref{fig:gravitino_5eV_CMB}, 
adding the measurement of the Simons Array to the observation of Planck,
we see that 
there are strong improvements on sensitivities to 
$f_{3/2}$ and $N_{3/2}$
because the Simons Array can precisely observe the CMB polarizations,
which is quite useful for getting the information of CMB lensing.
%
COrE+ can further improve  the measurement of CMB polarization and hence give tighter constraints.
Moreover, adding BAO data to CMB observations,
we find that 
constraints on $N_{3/2}$ are improved somewhat 
because several parameter degeneracies are broken by those combinations.

From these results,
we find that we can 
detect the nonzero values of $f_{3/2}$ and $N_{3/2}$ at 2$\sigma$ level
by using combinations of next generation CMB observations with BAO data and $H_0$
if $f_{3/2}$, i.e. $m_{3/2}$ has a relatively large value 
($f_{3/2}=$0.05353, i.e. $m_{3/2}=5$~eV). 
However, it is difficult to obtain lower bounds of $N_{3/2}$
even by using COrE+
if $f_{3/2}$ has a relatively small value 
($f_{3/2}=$0.01071, i.e. $m_{3/2}=1$~eV).
Additionally,
from the lower panels of Figs.\ref{fig:gravitino_1eV}-\ref{fig:gravitino_5eV_CMB},
these constraints becomes weaker
when we treat the total neutrino mass as a free parameter
because there is a degeneracy between 
the effect of massive neutrino and that of light gravitino.
In that case,
we can obtain a lower bound of $f_{3/2}$ 
only by using the COrE+ experiment.

%
On the other hand, 
from Figs.\ref{fig:gravitino_1eV}-\ref{fig:gravitino_5eV},
adding the 21 cm experiments to the CMB observations, 
we see that there are substantial improvements.
In particular,
the combination of SKA phase~1 with Planck + Simons Array, DESI and $H_0$
has enough sensitivity to obtaining a lower bound of $f_{3/2}$ at 2~$\sigma$ level
even when the fiducial value of $f_{3/2}$ is 0.01071
and we treat the total neutrino mass as a free parameter.
Furthermore,
the combination of SKA phase~2 with Planck + Simons Array, DESI and $H_0$
can detect the nonzero value of $N_{3/2}$
except when we treat the total neutrino mass as a free parameter.
If we use the combination of SKA phase~2 with COrE+, DESI and $H_0$,
we can detect its nonzero value even in that case.
Of course, Omniscope has enough sensitivity
to detect the signature of light gravitino in any cases.

\begin{figure}[htbp]
  \centering
  \includegraphics[bb=51 291 563 502,width=1\linewidth]{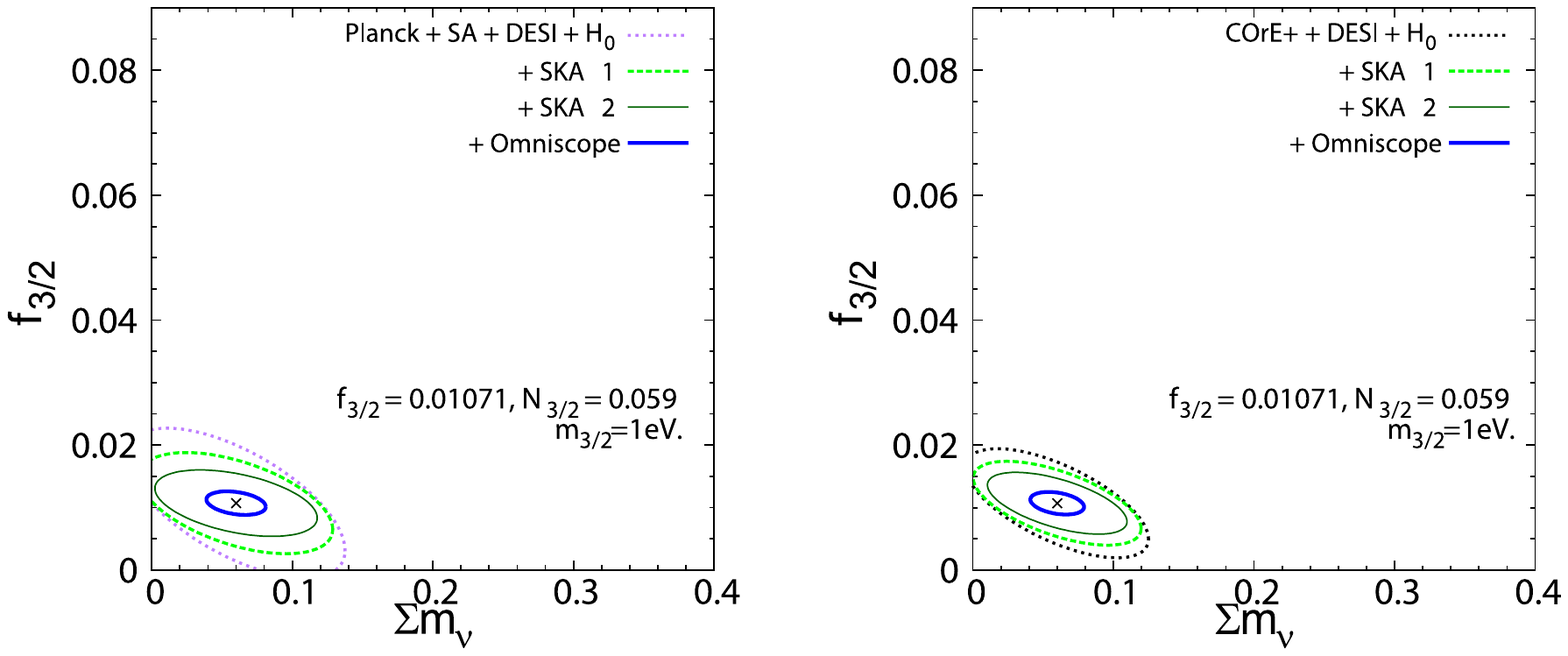}
  \includegraphics[bb=51 291 563 502,width=1\linewidth]{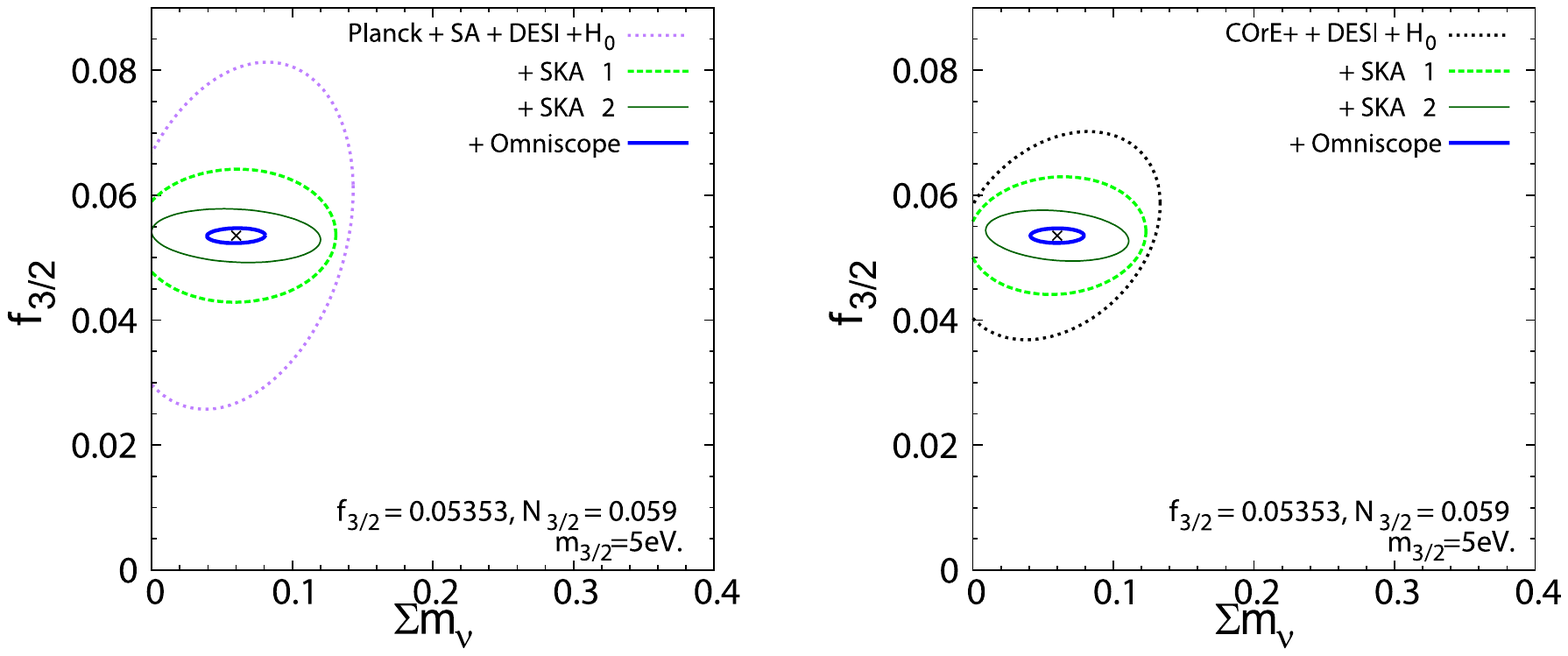}
  \caption{
Contours of 95\% C.L. forecasts in $\Sigma m_{\nu}$-$f_{3/2}$ plane.
We assume that $\Sigma m_{\nu} = 0.06$~eV and $N_{3/2}=0.059$.
In the upper panels, 
we assume $f_{3/2}=0.01071$, which correspond to $m_{3/2}=1$~eV
and in the lower panels, we assume $f_{3/2}=0.05353$, which correspond to $m_{3/2}=5$~eV.
We show constraints from Planck + Simons Array~(SA) + DESI~(BAO) + $H_{0}$ (dotted purple line)
with SKA phase 1 (dashed yellow-green line), phase 2 (solid green line) 
or Omniscope (thick blue line) in the left panels, 
and COrE+ + DESI~(BAO) + $H_{0}$ 
(dotted black line) with SKA phase 1, phase 2 or Omniscope in the right panels.
}\label{fig:gravitino_mnu}
\end{figure}

Moreover, in Figs.~\ref{fig:gravitino_mnu},
we plot contours of 95\% C.L. forecasts 
of each combination of CMB, 21cm line, DESI and $H_0$
in $\Sigma m_{\nu}$-$f_{3/2}$ plane. 
From the figure,
by using the combination of Planck +  Simons Array with DESI and $H_0$,
it is difficult to discriminate between
effects of massive neutrino and light gravitino
if the fiducial value of $f_{3/2}$ is $0.01071$.
%
%
However, even in that case,
we can discriminate them and obtain a lower bound of $f_{3/2}$
if we use the combination of SKA phase~1 with Planck + Simons Array, DESI and $H_0$.
Additionally, if we use the combination
of SKA phase~2 with COrE+, DESI and $H_0$
or Omniscope,
we can also detect the nonzero neutrino mass, simultaneously.

From our results, we find that 
21 cm line observations are
quite useful to constrain the mass of light gravitino,
and can significantly improve 
constraints on $f_{3/2}$ and $N_{3/2}$
in combination
with CMB, BAO and $H_0$ observations.
Besides, by using 21 cm line observations,
we 
will be able to 
discriminate between
effects of massive neutrino and light gravitino
through measuring the difference of 
their free streaming scales.


\begin{table}[htbp]
\centering \scalebox{0.75}[0.75]{
\begin{tabular}{l|ccccc}
\hline \hline 
& $\sigma( \Omega_{m}h^{2} )$ & $\sigma( \Omega_{b}h^{2} )$ &  $\sigma( \Omega_{\Lambda} )$ & $ \sigma( n_s ) $ & $\sigma( A_{s} \times 10^{10} )$ 
\\
\hline
Planck 
 & $ 1.32 \times 10^{  -3} $ & $ 2.07 \times 10^{  -4} $ & $ 9.54 \times 10^{  -3} $ & $ 7.16 \times 10^{  -3} $ & $ 1.92 \times 10^{  -1} $ 
\\
+ Simons Array (SA)
 & $ 5.95 \times 10^{  -4} $ & $ 6.85 \times 10^{  -5} $ & $ 3.87 \times 10^{  -3} $ & $ 2.95 \times 10^{  -3} $ & $ 1.46 \times 10^{  -1} $ 
\\
+ SA + BAO + $H_{0}$
 & $ 4.63 \times 10^{  -4} $ & $ 6.28 \times 10^{  -5} $ & $ 2.79 \times 10^{  -3} $ & $ 2.65 \times 10^{  -3} $ & $ 1.42 \times 10^{  -1} $ 
\\
+ SA + BAO + $H_{0}$ + SKA1
 & $ 4.17 \times 10^{  -4} $ & $ 6.09 \times 10^{  -5} $ & $ 2.50 \times 10^{  -3} $ & $ 2.57 \times 10^{  -3} $ & $ 1.35 \times 10^{  -1} $ 
\\
+ SA + BAO + $H_{0}$ + SKA2
 & $ 3.31 \times 10^{  -4} $ & $ 5.82 \times 10^{  -5} $ & $ 1.88 \times 10^{  -3} $ & $ 2.32 \times 10^{  -3} $ & $ 1.24 \times 10^{  -1} $ 
\\
+ SA + BAO + $H_{0}$ + Omniscope
 & $ 6.26 \times 10^{  -5} $ & $ 1.40 \times 10^{  -5} $ & $ 4.68 \times 10^{  -4} $ & $ 1.43 \times 10^{  -3} $ & $ 1.04 \times 10^{  -1} $ 
\\
COrE+
 & $ 4.88 \times 10^{  -4} $ & $ 5.20 \times 10^{  -5} $ & $ 3.08 \times 10^{  -3} $ & $ 2.47 \times 10^{  -3} $ & $ 8.33 \times 10^{  -2} $ 
\\
+ BAO + $H_{0}$
 & $ 4.08 \times 10^{  -4} $ & $ 4.97 \times 10^{  -5} $ & $ 2.46 \times 10^{  -3} $ & $ 2.35 \times 10^{  -3} $ & $ 8.27 \times 10^{  -2} $ 
\\
+ BAO + $H_{0}$ + SKA1
 & $ 3.70 \times 10^{  -4} $ & $ 4.87 \times 10^{  -5} $ & $ 2.23 \times 10^{  -3} $ & $ 2.28 \times 10^{  -3} $ & $ 8.11 \times 10^{  -2} $ 
\\
+ BAO + $H_{0}$ + SKA2
 & $ 2.91 \times 10^{  -4} $ & $ 4.73 \times 10^{  -5} $ & $ 1.66 \times 10^{  -3} $ & $ 2.05 \times 10^{  -3} $ & $ 7.67 \times 10^{  -2} $ 
\\
+ BAO + $H_{0}$ + Omniscope
 & $ 6.14 \times 10^{  -5} $ & $ 1.37 \times 10^{  -5} $ & $ 4.57 \times 10^{  -4} $ & $ 1.36 \times 10^{  -3} $ & $ 6.12 \times 10^{  -2} $ 
\\
\hline
& $\sigma( \tau )$  & $\sigma( Y_{p} )$ & $\sigma( f_{3/2} )$
\\
\hline
Planck 
 & $ 4.25 \times 10^{  -3} $ & $ 1.13 \times 10^{  -2} $ & $ 1.86 \times 10^{  -2} $
\\
+ Simons Array (SA)
 & $ 3.72 \times 10^{  -3} $ & $ 3.10 \times 10^{  -3} $ & $ 3.95 \times 10^{  -3} $
\\
+ SA + BAO + $H_{0}$
 & $ 3.57 \times 10^{  -3} $ & $ 2.96 \times 10^{  -3} $ & $ 3.46 \times 10^{  -3} $
\\
+ SA + BAO + $H_{0}$ + SKA1
 & $ 3.41 \times 10^{  -3} $ & $ 2.89 \times 10^{  -3} $ & $ 2.63 \times 10^{  -3} $ 
\\
+ SA + BAO + $H_{0}$ + SKA2
 & $ 3.10 \times 10^{  -3} $ & $ 2.75 \times 10^{  -3} $ & $ 1.65 \times 10^{  -3} $ 
\\
+ SA + BAO + $H_{0}$ + Omniscope
 & $ 2.47 \times 10^{  -3} $ & $ 1.31 \times 10^{  -3} $ & $ 7.13 \times 10^{  -4} $ 
\\
COrE+
 & $ 2.12 \times 10^{  -3} $ & $ 2.48 \times 10^{  -3} $ & $ 2.69 \times 10^{  -3} $
\\
+ BAO + $H_{0}$
 & $ 2.07 \times 10^{  -3} $ & $ 2.43 \times 10^{  -3} $ & $ 2.36 \times 10^{  -3} $
\\
+ BAO + $H_{0}$ + SKA1
 & $ 2.04 \times 10^{  -3} $ & $ 2.39 \times 10^{  -3} $ & $ 2.06 \times 10^{  -3} $ 
\\
+ BAO + $H_{0}$ + SKA2
 & $ 1.94 \times 10^{  -3} $ & $ 2.29 \times 10^{  -3} $ & $ 1.47 \times 10^{  -3} $ 
\\
+ BAO + $H_{0}$ + Omniscope
 & $ 1.48 \times 10^{  -3} $ & $ 1.16 \times 10^{  -3} $ & $ 6.78 \times 10^{  -4} $ 
\\
\hline
\end{tabular} }
\caption{1~$\sigma$ errors on cosmological parameters for fiducial $f_{3/2}=0.01071$ ($m_{3/2}=1$~eV) for the cases with fixed $\Sigma m_{\nu}=0.06$~eV and $N_{3/2}=0.059$.}
\label{tab:gravi1eV_fixN32}

\vspace{10pt}

\centering \scalebox{0.75}[0.75]{
\begin{tabular}{l|ccccc}
\hline \hline 
& $\sigma( \Omega_{m}h^{2} )$ & $\sigma( \Omega_{b}h^{2} )$ &  $\sigma( \Omega_{\Lambda} )$ & $ \sigma( n_s ) $ & $\sigma( A_{s} \times 10^{10} )$ 
\\
\hline
Planck 
 & $ 1.19 \times 10^{  -3} $ & $ 2.03 \times 10^{  -4} $ & $ 8.08 \times 10^{  -3} $ & $ 6.81 \times 10^{  -3} $ & $ 1.92 \times 10^{  -1} $ 
\\
+ Simons Array (SA)
 & $ 4.57 \times 10^{  -4} $ & $ 6.84 \times 10^{  -5} $ & $ 2.85 \times 10^{  -3} $ & $ 2.86 \times 10^{  -3} $ & $ 1.43 \times 10^{  -1} $ 
\\
+ SA + BAO + $H_{0}$
 & $ 4.01 \times 10^{  -4} $ & $ 6.25 \times 10^{  -5} $ & $ 2.33 \times 10^{  -3} $ & $ 2.65 \times 10^{  -3} $ & $ 1.33 \times 10^{  -1} $ 
\\
+ SA + BAO + $H_{0}$ + SKA1
 & $ 3.71 \times 10^{  -4} $ & $ 6.06 \times 10^{  -5} $ & $ 2.19 \times 10^{  -3} $ & $ 2.62 \times 10^{  -3} $ & $ 1.30 \times 10^{  -1} $ 
\\
+ SA + BAO + $H_{0}$ + SKA2
 & $ 3.07 \times 10^{  -4} $ & $ 5.78 \times 10^{  -5} $ & $ 1.79 \times 10^{  -3} $ & $ 2.51 \times 10^{  -3} $ & $ 1.22 \times 10^{  -1} $ 
\\
+ SA + BAO + $H_{0}$ + Omniscope
 & $ 5.22 \times 10^{  -5} $ & $ 1.20 \times 10^{  -5} $ & $ 4.11 \times 10^{  -4} $ & $ 9.73 \times 10^{  -4} $ & $ 1.05 \times 10^{  -1} $ 
\\
COrE+
 & $ 3.25 \times 10^{  -4} $ & $ 5.19 \times 10^{  -5} $ & $ 1.96 \times 10^{  -3} $ & $ 2.43 \times 10^{  -3} $ & $ 8.23 \times 10^{  -2} $ 
\\
+ BAO + $H_{0}$
 & $ 3.05 \times 10^{  -4} $ & $ 4.97 \times 10^{  -5} $ & $ 1.77 \times 10^{  -3} $ & $ 2.36 \times 10^{  -3} $ & $ 7.99 \times 10^{  -2} $ 
\\
+ BAO + $H_{0}$ + SKA1
 & $ 2.78 \times 10^{  -4} $ & $ 4.87 \times 10^{  -5} $ & $ 1.63 \times 10^{  -3} $ & $ 2.34 \times 10^{  -3} $ & $ 7.87 \times 10^{  -2} $ 
\\
+ BAO + $H_{0}$ + SKA2
 & $ 2.42 \times 10^{  -4} $ & $ 4.72 \times 10^{  -5} $ & $ 1.40 \times 10^{  -3} $ & $ 2.28 \times 10^{  -3} $ & $ 7.52 \times 10^{  -2} $ 
\\
+ BAO + $H_{0}$ + Omniscope
 & $ 5.10 \times 10^{  -5} $ & $ 1.18 \times 10^{  -5} $ & $ 3.95 \times 10^{  -4} $ & $ 9.56 \times 10^{  -4} $ & $ 6.18 \times 10^{  -2} $ 
\\
\hline
& $\sigma( \tau )$  & $\sigma( Y_{p} )$ & $\sigma( f_{3/2} )$
\\
\hline
Planck 
 & $ 4.28 \times 10^{  -3} $ & $ 1.13 \times 10^{  -2} $ & $ 5.73 \times 10^{  -2} $
\\
+ Simons Array (SA)
 & $ 3.71 \times 10^{  -3} $ & $ 3.12 \times 10^{  -3} $ & $ 8.97 \times 10^{  -3} $
\\
+ SA + BAO + $H_{0}$
 & $ 3.43 \times 10^{  -3} $ & $ 2.94 \times 10^{  -3} $ & $ 8.88 \times 10^{  -3} $
\\
+ SA + BAO + $H_{0}$ + SKA1
 & $ 3.33 \times 10^{  -3} $ & $ 2.85 \times 10^{  -3} $ & $ 4.15 \times 10^{  -3} $ 
\\
+ SA + BAO + $H_{0}$ + SKA2
 & $ 3.08 \times 10^{  -3} $ & $ 2.74 \times 10^{  -3} $ & $ 1.64 \times 10^{  -3} $ 
\\
+ SA + BAO + $H_{0}$ + Omniscope
 & $ 2.50 \times 10^{  -3} $ & $ 1.12 \times 10^{  -3} $ & $ 4.37 \times 10^{  -4} $ 
\\
COrE+
 & $ 2.13 \times 10^{  -3} $ & $ 2.51 \times 10^{  -3} $ & $ 5.30 \times 10^{  -3} $
\\
+ BAO + $H_{0}$
 & $ 2.06 \times 10^{  -3} $ & $ 2.43 \times 10^{  -3} $ & $ 5.22 \times 10^{  -3} $
\\
+ BAO + $H_{0}$ + SKA1
 & $ 2.03 \times 10^{  -3} $ & $ 2.37 \times 10^{  -3} $ & $ 3.52 \times 10^{  -3} $ 
\\
+ BAO + $H_{0}$ + SKA2
 & $ 1.93 \times 10^{  -3} $ & $ 2.31 \times 10^{  -3} $ & $ 1.57 \times 10^{  -3} $ 
\\
+ BAO + $H_{0}$ + Omniscope
 & $ 1.47 \times 10^{  -3} $ & $ 9.93 \times 10^{  -4} $ & $ 4.32 \times 10^{  -4} $ 
\\
\hline
\end{tabular} }
\caption{Same as in Table \ref{tab:gravi1eV_fixN32} but for fiducial $f_{3/2}=0.05353$ ($m_{3/2}=5$~eV).}
\label{tab:gravi5eV_fixN32}

\end{table}

\begin{table}[htbp]
\centering \scalebox{0.75}[0.75]{
\begin{tabular}{l|ccccc}
\hline \hline 
& $\sigma( \Omega_{m}h^{2} )$ & $\sigma( \Omega_{b}h^{2} )$ &  $\sigma( \Omega_{\Lambda} )$ & $ \sigma( n_s ) $ & $\sigma( A_{s} \times 10^{10} )$ 
\\
\hline
Planck 
 & $ 5.44 \times 10^{  -3} $ & $ 2.12 \times 10^{  -4} $ & $ 2.04 \times 10^{  -2} $ & $ 7.37 \times 10^{  -3} $ & $ 2.06 \times 10^{  -1} $ 
\\
+ Simons Array (SA)
 & $ 2.06 \times 10^{  -3} $ & $ 6.86 \times 10^{  -5} $ & $ 7.14 \times 10^{  -3} $ & $ 2.97 \times 10^{  -3} $ & $ 1.71 \times 10^{  -1} $ 
\\
+ SA + BAO + $H_{0}$
 & $ 1.42 \times 10^{  -3} $ & $ 6.32 \times 10^{  -5} $ & $ 4.93 \times 10^{  -3} $ & $ 2.65 \times 10^{  -3} $ & $ 1.57 \times 10^{  -1} $ 
\\
+ SA + BAO + $H_{0}$ + SKA1
 & $ 8.16 \times 10^{  -4} $ & $ 6.22 \times 10^{  -5} $ & $ 3.21 \times 10^{  -3} $ & $ 2.57 \times 10^{  -3} $ & $ 1.46 \times 10^{  -1} $ 
\\
+ SA + BAO + $H_{0}$ + SKA2
 & $ 5.16 \times 10^{  -4} $ & $ 5.95 \times 10^{  -5} $ & $ 2.13 \times 10^{  -3} $ & $ 2.32 \times 10^{  -3} $ & $ 1.39 \times 10^{  -1} $ 
\\
+ SA + BAO + $H_{0}$ + Omniscope
 & $ 6.26 \times 10^{  -5} $ & $ 2.27 \times 10^{  -5} $ & $ 4.92 \times 10^{  -4} $ & $ 1.58 \times 10^{  -3} $ & $ 1.06 \times 10^{  -1} $ 
\\
COrE+
 & $ 1.77 \times 10^{  -3} $ & $ 5.21 \times 10^{  -5} $ & $ 6.19 \times 10^{  -3} $ & $ 2.47 \times 10^{  -3} $ & $ 1.00 \times 10^{  -1} $ 
\\
+ BAO + $H_{0}$
 & $ 1.31 \times 10^{  -3} $ & $ 4.98 \times 10^{  -5} $ & $ 4.58 \times 10^{  -3} $ & $ 2.35 \times 10^{  -3} $ & $ 9.32 \times 10^{  -2} $ 
\\
+ BAO + $H_{0}$ + SKA1
 & $ 7.93 \times 10^{  -4} $ & $ 4.94 \times 10^{  -5} $ & $ 3.06 \times 10^{  -3} $ & $ 2.29 \times 10^{  -3} $ & $ 8.70 \times 10^{  -2} $ 
\\
+ BAO + $H_{0}$ + SKA2
 & $ 5.05 \times 10^{  -4} $ & $ 4.81 \times 10^{  -5} $ & $ 2.06 \times 10^{  -3} $ & $ 2.08 \times 10^{  -3} $ & $ 8.44 \times 10^{  -2} $ 
\\
+ BAO + $H_{0}$ + Omniscope
 & $ 6.14 \times 10^{  -5} $ & $ 2.11 \times 10^{  -5} $ & $ 4.81 \times 10^{  -4} $ & $ 1.50 \times 10^{  -3} $ & $ 6.42 \times 10^{  -2} $ 
\\
\hline
& $\sigma( \tau )$  & $\sigma( Y_{p} )$ & $\sigma( f_{3/2} )$ & $\sigma( N_{3/2} )$ 
\\
\hline
Planck 
 & $ 4.27 \times 10^{  -3} $ & $ 1.59 \times 10^{  -2} $ & $ 2.34 \times 10^{  -2} $ & $ 1.93 \times 10^{  -1} $
\\
+ Simons Array (SA)
 & $ 3.83 \times 10^{  -3} $ & $ 4.43 \times 10^{  -3} $ & $ 4.52 \times 10^{  -3} $ & $ 7.19 \times 10^{  -2} $
\\
+ SA + BAO + $H_{0}$
 & $ 3.65 \times 10^{  -3} $ & $ 3.83 \times 10^{  -3} $ & $ 3.75 \times 10^{  -3} $ & $ 5.07 \times 10^{  -2} $
\\
+ SA + BAO + $H_{0}$ + SKA1
 & $ 3.55 \times 10^{  -3} $ & $ 3.42 \times 10^{  -3} $ & $ 2.79 \times 10^{  -3} $ & $ 2.98 \times 10^{  -2} $ 
\\
+ SA + BAO + $H_{0}$ + SKA2
 & $ 3.39 \times 10^{  -3} $ & $ 3.17 \times 10^{  -3} $ & $ 1.92 \times 10^{  -3} $ & $ 2.07 \times 10^{  -2} $ 
\\
+ SA + BAO + $H_{0}$ + Omniscope
 & $ 2.57 \times 10^{  -3} $ & $ 2.06 \times 10^{  -3} $ & $ 7.23 \times 10^{  -4} $ & $ 6.43 \times 10^{  -3} $ 
\\
COrE+
 & $ 2.13 \times 10^{  -3} $ & $ 3.63 \times 10^{  -3} $ & $ 3.26 \times 10^{  -3} $ & $ 5.89 \times 10^{  -2} $
\\
+ BAO + $H_{0}$
 & $ 2.08 \times 10^{  -3} $ & $ 3.22 \times 10^{  -3} $ & $ 2.66 \times 10^{  -3} $ & $ 4.48 \times 10^{  -2} $
\\
+ BAO + $H_{0}$ + SKA1
 & $ 2.07 \times 10^{  -3} $ & $ 2.86 \times 10^{  -3} $ & $ 2.16 \times 10^{  -3} $ & $ 2.73 \times 10^{  -2} $ 
\\
+ BAO + $H_{0}$ + SKA2
 & $ 2.03 \times 10^{  -3} $ & $ 2.65 \times 10^{  -3} $ & $ 1.61 \times 10^{  -3} $ & $ 1.77 \times 10^{  -2} $ 
\\
+ BAO + $H_{0}$ + Omniscope
 & $ 1.58 \times 10^{  -3} $ & $ 1.80 \times 10^{  -3} $ & $ 6.88 \times 10^{  -4} $ & $ 5.80 \times 10^{  -3} $ 
\\
\hline
\end{tabular} }
\caption{1~$\sigma$ errors on cosmological parameters for fiducial $f_{3/2}=0.01071$ and $N_{3/2}=0.059$ ($m_{3/2}=1$~eV) for the cases with fixed $\Sigma m_{\nu}=0.06$~eV.}
\label{tab:gravi1eV}

\vspace{10pt}

\centering \scalebox{0.75}[0.75]{
\begin{tabular}{l|ccccc}
\hline \hline
& $\sigma( \Omega_{m}h^{2} )$ & $\sigma( \Omega_{b}h^{2} )$ &  $\sigma( \Omega_{\Lambda} )$ & $ \sigma( n_s ) $ & $\sigma( A_{s} \times 10^{10} )$ 
\\
\hline
Planck 
 & $ 2.98 \times 10^{  -3} $ & $ 2.09 \times 10^{  -4} $ & $ 1.40 \times 10^{  -2} $ & $ 7.19 \times 10^{  -3} $ & $ 1.97 \times 10^{  -1} $ 
\\
+ Simons Array (SA)
 & $ 1.02 \times 10^{  -3} $ & $ 6.90 \times 10^{  -5} $ & $ 5.26 \times 10^{  -3} $ & $ 3.07 \times 10^{  -3} $ & $ 1.53 \times 10^{  -1} $ 
\\
+ SA + BAO + $H_{0}$
 & $ 7.48 \times 10^{  -4} $ & $ 6.27 \times 10^{  -5} $ & $ 3.64 \times 10^{  -3} $ & $ 2.68 \times 10^{  -3} $ & $ 1.51 \times 10^{  -1} $ 
\\
+ SA + BAO + $H_{0}$ + SKA1
 & $ 5.48 \times 10^{  -4} $ & $ 6.06 \times 10^{  -5} $ & $ 2.88 \times 10^{  -3} $ & $ 2.64 \times 10^{  -3} $ & $ 1.42 \times 10^{  -1} $ 
\\
+ SA + BAO + $H_{0}$ + SKA2
 & $ 3.95 \times 10^{  -4} $ & $ 5.79 \times 10^{  -5} $ & $ 2.14 \times 10^{  -3} $ & $ 2.52 \times 10^{  -3} $ & $ 1.33 \times 10^{  -1} $ 
\\
+ SA + BAO + $H_{0}$ + Omniscope
 & $ 5.38 \times 10^{  -5} $ & $ 1.48 \times 10^{  -5} $ & $ 4.28 \times 10^{  -4} $ & $ 1.07 \times 10^{  -3} $ & $ 1.06 \times 10^{  -1} $ 
\\
COrE+
 & $ 8.56 \times 10^{  -4} $ & $ 5.26 \times 10^{  -5} $ & $ 4.44 \times 10^{  -3} $ & $ 2.67 \times 10^{  -3} $ & $ 8.72 \times 10^{  -2} $ 
\\
+ BAO + $H_{0}$
 & $ 6.63 \times 10^{  -4} $ & $ 4.97 \times 10^{  -5} $ & $ 3.32 \times 10^{  -3} $ & $ 2.43 \times 10^{  -3} $ & $ 8.70 \times 10^{  -2} $ 
\\
+ BAO + $H_{0}$ + SKA1
 & $ 5.23 \times 10^{  -4} $ & $ 4.88 \times 10^{  -5} $ & $ 2.72 \times 10^{  -3} $ & $ 2.38 \times 10^{  -3} $ & $ 8.46 \times 10^{  -2} $ 
\\
+ BAO + $H_{0}$ + SKA2
 & $ 3.81 \times 10^{  -4} $ & $ 4.74 \times 10^{  -5} $ & $ 2.04 \times 10^{  -3} $ & $ 2.30 \times 10^{  -3} $ & $ 8.18 \times 10^{  -2} $ 
\\
+ BAO + $H_{0}$ + Omniscope
 & $ 5.29 \times 10^{  -5} $ & $ 1.45 \times 10^{  -5} $ & $ 4.15 \times 10^{  -4} $ & $ 1.04 \times 10^{  -3} $ & $ 6.22 \times 10^{  -2} $ 
\\
\hline
& $\sigma( \tau )$  & $\sigma( Y_{p} )$ & $\sigma( f_{3/2} )$ & $\sigma( N_{3/2} )$ 
\\
\hline
Planck 
 & $ 4.29 \times 10^{  -3} $ & $ 1.14 \times 10^{  -2} $ & $ 6.47 \times 10^{  -2} $ & $ 6.33 \times 10^{  -2} $
\\
+ Simons Array (SA)
 & $ 3.79 \times 10^{  -3} $ & $ 3.29 \times 10^{  -3} $ & $ 1.12 \times 10^{  -2} $ & $ 1.71 \times 10^{  -2} $
\\
+ SA + BAO + $H_{0}$
 & $ 3.70 \times 10^{  -3} $ & $ 2.97 \times 10^{  -3} $ & $ 1.08 \times 10^{  -2} $ & $ 1.42 \times 10^{  -2} $
\\
+ SA + BAO + $H_{0}$ + SKA1
 & $ 3.54 \times 10^{  -3} $ & $ 2.87 \times 10^{  -3} $ & $ 4.29 \times 10^{  -3} $ & $ 9.33 \times 10^{  -3} $ 
\\
+ SA + BAO + $H_{0}$ + SKA2
 & $ 3.29 \times 10^{  -3} $ & $ 2.76 \times 10^{  -3} $ & $ 1.72 \times 10^{  -3} $ & $ 6.65 \times 10^{  -3} $ 
\\
+ SA + BAO + $H_{0}$ + Omniscope
 & $ 2.52 \times 10^{  -3} $ & $ 1.12 \times 10^{  -3} $ & $ 4.83 \times 10^{  -4} $ & $ 9.10 \times 10^{  -4} $ 
\\
COrE+
 & $ 2.14 \times 10^{  -3} $ & $ 2.62 \times 10^{  -3} $ & $ 6.57 \times 10^{  -3} $ & $ 1.26 \times 10^{  -2} $
\\
+ BAO + $H_{0}$
 & $ 2.10 \times 10^{  -3} $ & $ 2.45 \times 10^{  -3} $ & $ 6.40 \times 10^{  -3} $ & $ 1.04 \times 10^{  -2} $
\\
+ BAO + $H_{0}$ + SKA1
 & $ 2.08 \times 10^{  -3} $ & $ 2.39 \times 10^{  -3} $ & $ 3.79 \times 10^{  -3} $ & $ 7.62 \times 10^{  -3} $ 
\\
+ BAO + $H_{0}$ + SKA2
 & $ 2.01 \times 10^{  -3} $ & $ 2.32 \times 10^{  -3} $ & $ 1.60 \times 10^{  -3} $ & $ 5.29 \times 10^{  -3} $ 
\\
+ BAO + $H_{0}$ + Omniscope
 & $ 1.50 \times 10^{  -3} $ & $ 9.93 \times 10^{  -4} $ & $ 4.73 \times 10^{  -4} $ & $ 8.93 \times 10^{  -4} $ 
\\
\hline
\end{tabular} }
\caption{Same as in Table \ref{tab:gravi1eV} but for fiducial $f_{3/2}=0.05353$ ($m_{3/2}=5$~eV).}
\label{tab:gravi5eV}

\end{table}

\begin{table}[htbp]
\centering \scalebox{0.75}[0.75]{
\begin{tabular}{l|ccccc}
\hline \hline 
& $\sigma( \Omega_{m}h^{2} )$ & $\sigma( \Omega_{b}h^{2} )$ &  $\sigma( \Omega_{\Lambda} )$ & $ \sigma( n_s ) $ & $\sigma( A_{s} \times 10^{10} )$ 
\\
\hline
Planck 
 & $ 5.51 \times 10^{  -3} $ & $ 2.35 \times 10^{  -4} $ & $ 2.47 \times 10^{  -2} $ & $ 7.61 \times 10^{  -3} $ & $ 2.07 \times 10^{  -1} $ 
\\
+ Simons Array (SA)
 & $ 2.09 \times 10^{  -3} $ & $ 7.27 \times 10^{  -5} $ & $ 9.53 \times 10^{  -3} $ & $ 3.39 \times 10^{  -3} $ & $ 1.77 \times 10^{  -1} $ 
\\
+ SA + BAO + $H_{0}$
 & $ 1.43 \times 10^{  -3} $ & $ 6.50 \times 10^{  -5} $ & $ 5.05 \times 10^{  -3} $ & $ 2.72 \times 10^{  -3} $ & $ 1.68 \times 10^{  -1} $ 
\\
+ SA + BAO + $H_{0}$ + SKA1
 & $ 8.18 \times 10^{  -4} $ & $ 6.43 \times 10^{  -5} $ & $ 3.61 \times 10^{  -3} $ & $ 2.60 \times 10^{  -3} $ & $ 1.63 \times 10^{  -1} $ 
\\
+ SA + BAO + $H_{0}$ + SKA2
 & $ 5.19 \times 10^{  -4} $ & $ 6.31 \times 10^{  -5} $ & $ 2.37 \times 10^{  -3} $ & $ 2.34 \times 10^{  -3} $ & $ 1.58 \times 10^{  -1} $ 
\\
+ SA + BAO + $H_{0}$ + Omniscope
 & $ 7.78 \times 10^{  -5} $ & $ 2.29 \times 10^{  -5} $ & $ 6.65 \times 10^{  -4} $ & $ 1.60 \times 10^{  -3} $ & $ 1.16 \times 10^{  -1} $ 
\\
COrE+
 & $ 1.80 \times 10^{  -3} $ & $ 5.50 \times 10^{  -5} $ & $ 8.27 \times 10^{  -3} $ & $ 3.03 \times 10^{  -3} $ & $ 1.00 \times 10^{  -1} $ 
\\
+ BAO + $H_{0}$
 & $ 1.32 \times 10^{  -3} $ & $ 5.04 \times 10^{  -5} $ & $ 4.72 \times 10^{  -3} $ & $ 2.35 \times 10^{  -3} $ & $ 9.50 \times 10^{  -2} $ 
\\
+ BAO + $H_{0}$ + SKA1
 & $ 7.94 \times 10^{  -4} $ & $ 5.00 \times 10^{  -5} $ & $ 3.51 \times 10^{  -3} $ & $ 2.29 \times 10^{  -3} $ & $ 9.08 \times 10^{  -2} $ 
\\
+ BAO + $H_{0}$ + SKA2
 & $ 5.10 \times 10^{  -4} $ & $ 4.94 \times 10^{  -5} $ & $ 2.34 \times 10^{  -3} $ & $ 2.08 \times 10^{  -3} $ & $ 8.87 \times 10^{  -2} $ 
\\
+ BAO + $H_{0}$ + Omniscope
 & $ 7.46 \times 10^{  -5} $ & $ 2.15 \times 10^{  -5} $ & $ 6.50 \times 10^{  -4} $ & $ 1.51 \times 10^{  -3} $ & $ 7.30 \times 10^{  -2} $ 
\\
\hline%
& $\sigma( \tau )$  & $\sigma( Y_{p} )$ & $\sigma( f_{3/2} )$ & $\sigma( N_{3/2} )$ & $\sigma( \Sigma m_{\nu} )$ 
\\
\hline
Planck 
 & $ 4.29 \times 10^{  -3} $ & $ 1.66 \times 10^{  -2} $ & $ 2.82 \times 10^{  -2} $ & $ 2.38 \times 10^{  -1} $ & $ 2.02 \times 10^{  -1} $
\\
+ Simons Array (SA)
 & $ 4.07 \times 10^{  -3} $ & $ 4.85 \times 10^{  -3} $ & $ 6.63 \times 10^{  -3} $ & $ 9.55 \times 10^{  -2} $ & $ 8.44 \times 10^{  -2} $
\\
+ SA + BAO + $H_{0}$
 & $ 3.97 \times 10^{  -3} $ & $ 3.95 \times 10^{  -3} $ & $ 4.84 \times 10^{  -3} $ & $ 5.17 \times 10^{  -2} $ & $ 3.13 \times 10^{  -2} $
\\
+ SA + BAO + $H_{0}$ + SKA1
 & $ 3.95 \times 10^{  -3} $ & $ 3.52 \times 10^{  -3} $ & $ 3.26 \times 10^{  -3} $ & $ 3.04 \times 10^{  -2} $ & $ 2.77 \times 10^{  -2} $ 
\\
+ SA + BAO + $H_{0}$ + SKA2
 & $ 3.87 \times 10^{  -3} $ & $ 3.35 \times 10^{  -3} $ & $ 2.14 \times 10^{  -3} $ & $ 2.26 \times 10^{  -2} $ & $ 2.33 \times 10^{  -2} $ 
\\
+ SA + BAO + $H_{0}$ + Omniscope
 & $ 2.76 \times 10^{  -3} $ & $ 2.07 \times 10^{  -3} $ & $ 7.60 \times 10^{  -4} $ & $ 6.65 \times 10^{  -3} $ & $ 8.51 \times 10^{  -3} $ 
\\
COrE+
 & $ 2.16 \times 10^{  -3} $ & $ 4.11 \times 10^{  -3} $ & $ 5.04 \times 10^{  -3} $ & $ 8.30 \times 10^{  -2} $ & $ 7.04 \times 10^{  -2} $
\\
+ BAO + $H_{0}$
 & $ 2.15 \times 10^{  -3} $ & $ 3.33 \times 10^{  -3} $ & $ 3.52 \times 10^{  -3} $ & $ 4.71 \times 10^{  -2} $ & $ 2.62 \times 10^{  -2} $
\\
+ BAO + $H_{0}$ + SKA1
 & $ 2.15 \times 10^{  -3} $ & $ 2.92 \times 10^{  -3} $ & $ 2.71 \times 10^{  -3} $ & $ 2.86 \times 10^{  -2} $ & $ 2.41 \times 10^{  -2} $ 
\\
+ BAO + $H_{0}$ + SKA2
 & $ 2.13 \times 10^{  -3} $ & $ 2.78 \times 10^{  -3} $ & $ 1.99 \times 10^{  -3} $ & $ 2.11 \times 10^{  -2} $ & $ 2.00 \times 10^{  -2} $ 
\\
+ BAO + $H_{0}$ + Omniscope
 & $ 1.73 \times 10^{  -3} $ & $ 1.80 \times 10^{  -3} $ & $ 7.29 \times 10^{  -4} $ & $ 5.86 \times 10^{  -3} $ & $ 7.69 \times 10^{  -3} $ 
\\
\hline
\end{tabular} }
\caption{1~$\sigma$ errors on cosmological parameters for fiducial $f_{3/2}=0.01071$ and $N_{3/2}=0.059$ ($m_{3/2}=1$~eV) for the cases with  freely varying $\Sigma m_{\nu}$.}
\label{tab:gravi1eV_mnu_free}

\vspace{10pt}

\centering \scalebox{0.75}[0.75]{
\begin{tabular}{l|ccccc}
\hline \hline 
& $\sigma( \Omega_{m}h^{2} )$ & $\sigma( \Omega_{b}h^{2} )$ &  $\sigma( \Omega_{\Lambda} )$ & $ \sigma( n_s ) $ & $\sigma( A_{s} \times 10^{10} )$ 
\\
\hline
Planck 
 & $ 3.45 \times 10^{  -3} $ & $ 2.43 \times 10^{  -4} $ & $ 2.48 \times 10^{  -2} $ & $ 7.58 \times 10^{  -3} $ & $ 1.97 \times 10^{  -1} $ 
\\
+ Simons Array (SA)
 & $ 1.23 \times 10^{  -3} $ & $ 7.19 \times 10^{  -5} $ & $ 9.65 \times 10^{  -3} $ & $ 3.17 \times 10^{  -3} $ & $ 1.72 \times 10^{  -1} $ 
\\
+ SA + BAO + $H_{0}$
 & $ 7.74 \times 10^{  -4} $ & $ 6.48 \times 10^{  -5} $ & $ 3.71 \times 10^{  -3} $ & $ 2.83 \times 10^{  -3} $ & $ 1.60 \times 10^{  -1} $ 
\\
+ SA + BAO + $H_{0}$ + SKA1
 & $ 5.52 \times 10^{  -4} $ & $ 6.33 \times 10^{  -5} $ & $ 3.29 \times 10^{  -3} $ & $ 2.72 \times 10^{  -3} $ & $ 1.59 \times 10^{  -1} $ 
\\
+ SA + BAO + $H_{0}$ + SKA2
 & $ 3.96 \times 10^{  -4} $ & $ 6.19 \times 10^{  -5} $ & $ 2.41 \times 10^{  -3} $ & $ 2.66 \times 10^{  -3} $ & $ 1.56 \times 10^{  -1} $ 
\\
+ SA + BAO + $H_{0}$ + Omniscope
 & $ 7.60 \times 10^{  -5} $ & $ 1.61 \times 10^{  -5} $ & $ 6.84 \times 10^{  -4} $ & $ 1.13 \times 10^{  -3} $ & $ 1.13 \times 10^{  -1} $ 
\\
COrE+
 & $ 1.08 \times 10^{  -3} $ & $ 5.42 \times 10^{  -5} $ & $ 8.46 \times 10^{  -3} $ & $ 2.81 \times 10^{  -3} $ & $ 9.37 \times 10^{  -2} $ 
\\
+ BAO + $H_{0}$
 & $ 6.70 \times 10^{  -4} $ & $ 5.05 \times 10^{  -5} $ & $ 3.51 \times 10^{  -3} $ & $ 2.48 \times 10^{  -3} $ & $ 8.89 \times 10^{  -2} $ 
\\
+ BAO + $H_{0}$ + SKA1
 & $ 5.30 \times 10^{  -4} $ & $ 4.97 \times 10^{  -5} $ & $ 3.24 \times 10^{  -3} $ & $ 2.40 \times 10^{  -3} $ & $ 8.84 \times 10^{  -2} $ 
\\
+ BAO + $H_{0}$ + SKA2
 & $ 3.85 \times 10^{  -4} $ & $ 4.88 \times 10^{  -5} $ & $ 2.37 \times 10^{  -3} $ & $ 2.34 \times 10^{  -3} $ & $ 8.76 \times 10^{  -2} $ 
\\
+ BAO + $H_{0}$ + Omniscope
 & $ 7.37 \times 10^{  -5} $ & $ 1.56 \times 10^{  -5} $ & $ 6.64 \times 10^{  -4} $ & $ 1.08 \times 10^{  -3} $ & $ 6.95 \times 10^{  -2} $ 
\\
\hline
& $\sigma( \tau )$  & $\sigma( Y_{p} )$ & $\sigma( f_{3/2} )$ & $\sigma( N_{3/2} )$ & $\sigma( \Sigma m_{\nu} )$ 
\\
\hline
Planck 
 & $ 4.29 \times 10^{  -3} $ & $ 1.16 \times 10^{  -2} $ & $ 6.48 \times 10^{  -2} $ & $ 6.72 \times 10^{  -2} $ & $ 1.73 \times 10^{  -1} $
\\
+ Simons Array (SA)
 & $ 4.07 \times 10^{  -3} $ & $ 3.29 \times 10^{  -3} $ & $ 1.15 \times 10^{  -2} $ & $ 2.19 \times 10^{  -2} $ & $ 7.40 \times 10^{  -2} $
\\
+ SA + BAO + $H_{0}$
 & $ 3.96 \times 10^{  -3} $ & $ 3.24 \times 10^{  -3} $ & $ 1.12 \times 10^{  -2} $ & $ 1.92 \times 10^{  -2} $ & $ 3.36 \times 10^{  -2} $
\\
+ SA + BAO + $H_{0}$ + SKA1
 & $ 3.91 \times 10^{  -3} $ & $ 3.05 \times 10^{  -3} $ & $ 4.29 \times 10^{  -3} $ & $ 1.11 \times 10^{  -2} $ & $ 2.86 \times 10^{  -2} $ 
\\
+ SA + BAO + $H_{0}$ + SKA2
 & $ 3.85 \times 10^{  -3} $ & $ 2.97 \times 10^{  -3} $ & $ 1.73 \times 10^{  -3} $ & $ 7.75 \times 10^{  -3} $ & $ 2.42 \times 10^{  -2} $ 
\\
+ SA + BAO + $H_{0}$ + Omniscope
 & $ 2.66 \times 10^{  -3} $ & $ 1.25 \times 10^{  -3} $ & $ 4.84 \times 10^{  -4} $ & $ 9.30 \times 10^{  -4} $ & $ 8.33 \times 10^{  -3} $ 
\\
COrE+
 & $ 2.16 \times 10^{  -3} $ & $ 2.63 \times 10^{  -3} $ & $ 6.89 \times 10^{  -3} $ & $ 1.69 \times 10^{  -2} $ & $ 6.15 \times 10^{  -2} $
\\
+ BAO + $H_{0}$
 & $ 2.16 \times 10^{  -3} $ & $ 2.61 \times 10^{  -3} $ & $ 6.72 \times 10^{  -3} $ & $ 1.50 \times 10^{  -2} $ & $ 2.96 \times 10^{  -2} $
\\
+ BAO + $H_{0}$ + SKA1
 & $ 2.15 \times 10^{  -3} $ & $ 2.49 \times 10^{  -3} $ & $ 3.80 \times 10^{  -3} $ & $ 9.80 \times 10^{  -3} $ & $ 2.54 \times 10^{  -2} $ 
\\
+ BAO + $H_{0}$ + SKA2
 & $ 2.13 \times 10^{  -3} $ & $ 2.43 \times 10^{  -3} $ & $ 1.64 \times 10^{  -3} $ & $ 6.86 \times 10^{  -3} $ & $ 2.05 \times 10^{  -2} $ 
\\
+ BAO + $H_{0}$ + Omniscope
 & $ 1.64 \times 10^{  -3} $ & $ 1.09 \times 10^{  -3} $ & $ 4.73 \times 10^{  -4} $ & $ 9.19 \times 10^{  -4} $ & $ 7.66 \times 10^{  -3} $ 
\\
\hline
\end{tabular} }
\caption{Same as in Table \ref{tab:gravi1eV_mnu_free} but for fiducial $f_{3/2}=0.05353$ ($m_{3/2}=5$~eV).}
\label{tab:gravi5eV_mnu_free}

\end{table}

\section{Conclusion}
\label{sec:conclusion}

In this paper,
we have studied how well we can constrain the mass of light gravitino 
$m_{3/2}<\mathcal{O}(10)$~eV, 
or more specifically,
the fraction of light gravitinos in the total dark matter density $f_{3/2}$,
and the effective number of neutrino species 
for
light gravitinos $N_{3/2}$,
which determine $m_{3/2}$,
by using observations of 21 cm line, CMB, BAO and direct measurements of $H_0$.

In the early Universe,
light gravitinos are produced from thermal plasma,
and they behave as warm dark matter~(WDM) at late epochs.
Thus, they imprint characteristic signatures 
on density fluctuations,
and we can detect the features through cosmological observations,
such as CMB and 21 cm line.
Adding the measurement of the Simons Array,
which is a planned precise CMB polarization observation,
to the observation of Planck,
we see that there are strong improvements on sensitivities to constraints 
on $f_{3/2}$ and $N_{3/2}$
because the Simons Array is quite useful for getting the information of CMB lensing.
If $f_{3/2}$, i.e. $m_{3/2}$ has a relatively large value
($f_{3/2}=$0.05353, which corresponds to $m_{3/2}=5$~eV),
by using Planck + Simons Array or COrE+,
we can detect the nonzero values of $f_{3/2}$ and $N_{3/2}$ at 2$\sigma$ level.
%

%
Besides, adding the 21 cm experiments to the CMB observations, 
we see that there are substantial improvements.
For the cases  with fixed $N_{3/2}=0.059$ and $\Sigma m_{\nu}=0.06$~eV,
by combining SKA phase~1 with Planck, the Simons Array, DESI and 
a direct measurement of $H_0$ at 1\% accuracy,
we can obtain a 1~$\sigma$ error on the mass of light gravitinos,
$\sigma(m_{3/2})=0.25$~eV
for fiducial $f_{3/2}=0.01071$,
which corresponds to $m_{3/2}=1$~eV.
If we use SKA phase~2 or Omniscope,
the error can be improved as
$\sigma(m_{3/2})=0.16$~eV (SKA phase~2) or $\sigma(m_{3/2})=0.067$~eV (Omniscope), respectively. 
In particular,
the combination of SKA phase~1 with Planck + Simons Array, DESI and $H_0$
has enough sensitivity to obtaining a lower bound of $f_{3/2}$ at 2~$\sigma$ level
even when  the fiducial value of $f_{3/2}$ is %
as small as
0.01071
and we treat $N_{3/2}$ and the total neutrino mass as  free parameters.
Furthermore,
the combination of SKA phase~2 with Planck + Simons Array, DESI and $H_0$
can detect the nonzero value of $N_{3/2}$
except when we treat the total neutrino mass as a free parameter.
Moreover,
if we use the combination of SKA phase~2 with COrE+, DESI and $H_0$,
we can detect the nonzero value of $N_{3/2}$ even in that case.

Although 
it is difficult to discriminate between 
the effects of massive neutrinos and light gravitinos
only by using Planck + Simons Array, BAO and a measurement of $H_0$,
%
it becomes feasible if a precise observation of 21 cm line is incorporated.
%
In particular, 
the combination of SKA phase~2 with COrE+, DESI and $H_0$
has enough sensitivities to determine
the parameters of light gravitino and the total neutrino mass
at 2~$\sigma$ level, simultaneously.
If we use Omniscope, we can 
detect features of light gravitinos and massive neutrinos 
even with on-going CMB observations.

Our results indicate that 
combining 21 cm line observations with CMB observations
has strong impacts on the determination of the mass of light gravitinos
and understanding the origin of matter in the Universe.

\section*{Acknowledgments}

%

We thank Tomo Takahashi for a useful correspondence
about cosmological effects of light gravitino.
Besides, 
we appreciate the significant contribution made by
Toyokazu Sekiguchi.
This work is supported by MEXT KAKENHI Grant Number 15H05889 (M. K.),
JSPS KAKENHI Grant Number 25400248 (M. K.) and also by
the World Premier International Research Center Initiative (WPI), MEXT, Japan.

\end{document}